\documentclass[oneside,onecolumn]{article}

\usepackage{blindtext} 

\usepackage[sc]{mathpazo} 
\usepackage[T1]{fontenc} 
\usepackage{microtype} 

\usepackage[english]{babel} 

\usepackage[hmarginratio=1:1,top=32mm,columnsep=20pt]{geometry} 
\usepackage[hang, small,labelfont=bf,up,textfont=it,up]{caption} 
\usepackage{booktabs} 

\usepackage{lettrine} 

\usepackage{enumitem} 
\setlist[itemize]{noitemsep} 

\usepackage{abstract} 

\usepackage{titlesec} 
\renewcommand\thesection{\Roman{section}} 
\renewcommand\thesubsection{\roman{subsection}} 
\titleformat{\section}[block]{\large\scshape\centering}{\thesection.}{1em}{} 
\titleformat{\subsection}[block]{\large}{\thesubsection.}{1em}{} 

\usepackage{fancyhdr} 
\usepackage{titling} 

\usepackage{hyperref} 

\usepackage{scalerel, amssymb}

\def\msquare{\mathord{\scalerel*{\Box}{gX}}}
\usepackage{amsmath}
\usepackage{framed}
\usepackage{graphicx} 
\usepackage{hyperref}   
\usepackage{multicol}
\usepackage{float}
\usepackage{nomencl}
\usepackage{wasysym}
\hypersetup{
	colorlinks=true,
	linkcolor=blue,
	citecolor=blue,
	urlcolor=blue,
}
\usepackage[square,numbers]{natbib}

\pretitle{\begin{center}\huge\bfseries} 
\posttitle{\end{center}} 
\title{Identification of the dynamic force and moment characteristics of annular gaps using linear independent rotor whirling motions} 
\author{
\textsc{Maximilian M. G. Kuhr}\thanks{Corresponding author} \\[1ex]
\normalsize Chair of Fluid Systems \\
\normalsize Technische Universität Darmstadt \\
\normalsize \href{mailto:maximilian.kuhr@fst.tu-darmstadt.de}{maximilian.kuhr@fst.tu-darmstadt.de}
}
\date{} 
\renewcommand{\maketitlehookd}{%
\begin{abstract}
Nowadays, most studies on the dynamic properties of annular gaps focus only on the force characteristics due to translational motions, while the tilt and moment coefficients are less well studied. Therefore, there is hardly any reliable experimental data for the additional coefficients that can be used for validation purpose. To improve this, a test rig first presented by \cite{Kuhr.2022, Kuhr.2022b} is used to experimentally determine the dynamic force and moment characteristics of three annuli of different lengths. By using active magnetic bearings, the rotor is excited with user-defined frequencies and the rotor position and the forces and moments induced by the flow field in the annulus are measured. To obtain accurate and reliable experimental data, extensive preliminary studies are carried out to determine the known characteristics of the test rig rotor and the added mass and inertia imposed by the test rig. Subsequently, an elaborate uncertainty quantification is carried out to quantify the measurement uncertainties. The experimental results, i.e. the 48 rotordynamic coefficients, are compared to a calculation method by \citet{Kuhr.2022, Kuhr.2022c}. It is shown that the presented experimental data agree well with the calculation method, especially for the additional rotordynamic tilt and moment coefficients. Furthermore, it is shown that the annulus length significantly influences the coefficients of the first sub-matrix. A dependence of the additional coefficients on the length is recognisable, but less pronounced. Even for the shortest investigated annulus, i.e. $L := \tilde{L}/\tilde{R} = 1$, the stiffness coefficients due to the forces from the angular motion of the rotor are of the same order of magnitude as the stiffness coefficients due to the forces from the translational motion. This supports recent results by \citet{Kuhr.2022, Kuhr.2022c}, indicating that the additional coefficients become relevant much earlier than assumed throughout the literature, cf. \citet{Childs.1993}. 
\end{abstract}
}

\begin{document}

\maketitle


\section{Introduction}
Reliable and accurate experiments provide the foundation for the validation of any calculation model. By comparing the model to real components under well controlled conditions, conclusions can be drawn regarding the accuracy and reliability of the model itself. 
Regarding the dynamic force and moment characteristic of annular gaps, most experimental studies focus only on the forces due to translational motions, while the tilt and moment coefficients are less well studied. Therefore, there is hardly any reliable experimental data for the additional coefficients that can be used for validation purpose. 
This is of particular importance since the additional tilt and moment characteristics can have a significant impact on the rotordynamic behaviour of modern turbomachinery, i.e the eigenvalues and eigenfrequencies, cf. \citet{Gasch.2002, Feng.2017, Kim.2018}. Recent results by \citet{Kuhr.2022, Kuhr.2022c} for example show that the additional tilt and moment characteristics are becoming relevant much earlier than assumed throughout the literature. However, due to the limited experimental data, the results, which are solely based on simulation results, are hardly validated.

\begin{figure} 	
\centering 	
\includegraphics[width = \textwidth]{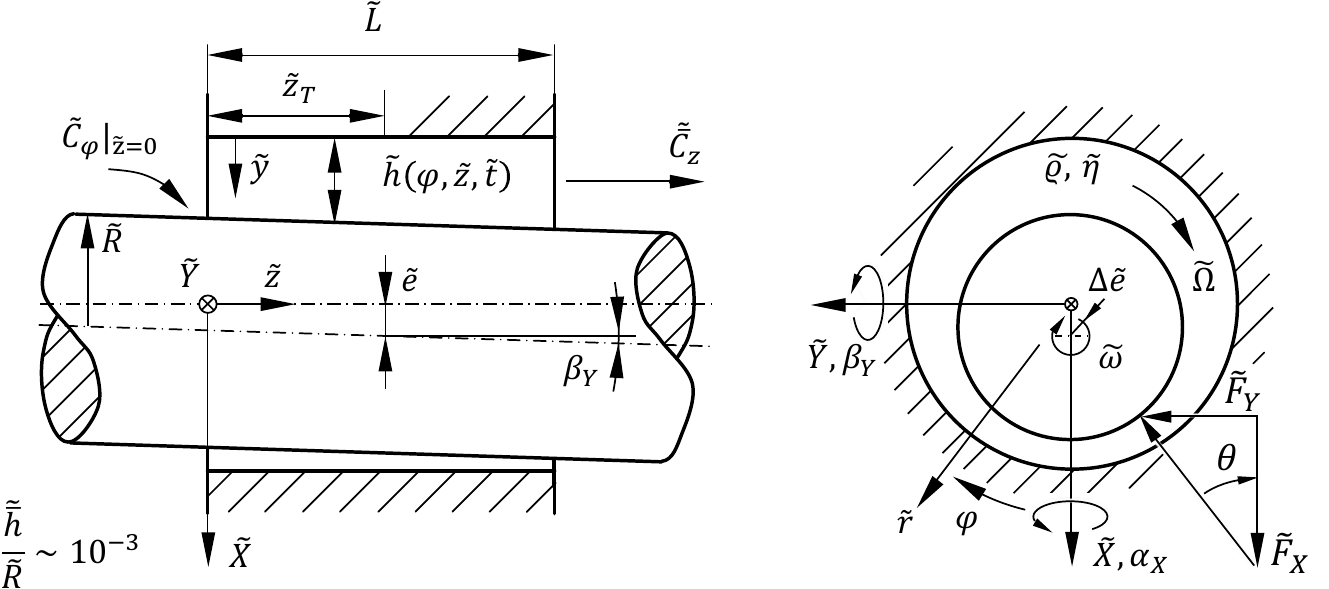}
	\caption{Schematic drawing of an eccentric operated generic annular gap with axial flow and pre-swirl.}
	\label{fig:generic_annulus}
\end{figure}

In modern turbomachinery, two essential narrow annular gaps exist, applying fluid-film forces and moments on the rotating shaft \citep{Childs.1993, Gasch.2002, Gulich.2010}. First, journal bearings which are either oil or media lubricated and second, annular seals or damper seals. The usage of low viscous fluids like water or cryogenic liquids for lubrication purpose, often leads to an operation at high Reynolds numbers, resulting in turbulent flow conditions and significant inertia effects \citep{Childs.1993, SanAndres.1991b, SanAndres.1993, SanAndres.1993b, SanAndres.1993c}. However, this paper neither focus on annular seals or journal bearings, but concentrate on a generic geometry valid for both machine elements, cf. figure \ref{fig:generic_annulus}.

In general, the dynamic characteristics of annular gaps are modelled by using classical mechanical elements, namely springs, dampers and masses. Accordingly, the system is characterised by using stiffness $\tilde{K}$, damping $\tilde{C}$ and inertia $\tilde{M}$ coefficients. Throughout the paper, the tilde $\tilde{\msquare}$ characterises dimensional variables. The generalised equation of motion, including forces and moments of the annular gap flow yields
\begin{subequations}\label{eqn:introduction_full_matrix}
\begin{gather}
    \begin{split}
		-\begin{bmatrix}
			\tilde{F}_{X}\\
			\tilde{F}_{Y}\\
			\tilde{M}_{X}\\
			\tilde{M}_{Y}
		\end{bmatrix} = \begin{bmatrix}
			\tilde{K}_{XX} &  \tilde{K}_{XY} & \tilde{K}_{X\alpha} & \tilde{K}_{X \beta}\\
			\tilde{K}_{YX} &  \tilde{K}_{YY} & \tilde{K}_{Y\alpha} & \tilde{K}_{Y \beta}\\
			\tilde{K}_{\alpha X} &  \tilde{K}_{\alpha Y} & \tilde{K}_{\alpha\alpha} & \tilde{K}_{\alpha\beta}\\
			\tilde{K}_{\beta X} &  \tilde{K}_{\beta Y} & \tilde{K}_{\beta\alpha} & \tilde{K}_{\beta\beta}
		\end{bmatrix} \begin{bmatrix} 
			\tilde{X}\\
			\tilde{Y}\\
			\alpha_X\\
			\beta_Y
		\end{bmatrix} + & \begin{bmatrix}
			\tilde{C}_{XX} & \tilde{C}_{XY} & \tilde{C}_{X\alpha} & \tilde{C}_{X \beta}\\
			\tilde{C}_{YX} & \tilde{C}_{YY} & \tilde{C}_{Y\alpha} & \tilde{C}_{Y \beta}\\
			\tilde{C}_{\alpha X} & \tilde{C}_{\alpha Y} & \tilde{C}_{\alpha\alpha} & \tilde{C}_{\alpha\beta}\\
			\tilde{C}_{\beta X} & \tilde{C}_{\beta Y} & \tilde{C}_{\beta\alpha} & \tilde{C}_{\beta\beta}
		\end{bmatrix} \begin{bmatrix} 
			\tilde{\dot{X}}\\
			\tilde{\dot{Y}}\\
			\tilde{\dot{\alpha}}_X\\
			\tilde{\dot{\beta}}_Y
		\end{bmatrix} + \\  &\qquad \qquad \qquad + \begin{bmatrix}
			\tilde{M}_{XX} & \tilde{M}_{XY} & \tilde{M}_{X\alpha} & \tilde{M}_{X \beta}\\
			\tilde{M}_{YX} & \tilde{M}_{YY} & \tilde{M}_{Y\alpha} & \tilde{M}_{Y \beta}\\
			\tilde{M}_{\alpha X} & \tilde{M}_{\alpha Y} & \tilde{M}_{\alpha\alpha} & \tilde{M}_{\alpha\beta}\\
			\tilde{M}_{\beta X} & \tilde{M}_{\beta Y} & \tilde{M}_{\beta\alpha} & \tilde{M}_{\beta\beta}
		\end{bmatrix} \begin{bmatrix} 
			\tilde{\ddot{X}}\\
			\tilde{\ddot{Y}}\\
			\tilde{\ddot{\alpha}}_X\\
			\tilde{\ddot{\beta}}_Y
		\end{bmatrix},	
	\end{split}\\
	\begin{split}
	    -\begin{bmatrix}
			\tilde{F}_{X}\\
			\tilde{F}_{Y}\\
			\tilde{M}_{X}\\
			\tilde{M}_{Y}
		\end{bmatrix} = &\begin{bmatrix}
			\tilde{\boldsymbol{K}}_{\mathrm{I}} & \tilde{\boldsymbol{K}}_{\mathrm{II}}\\
			\tilde{\boldsymbol{K}}_{\mathrm{III}} & \tilde{\boldsymbol{K}}_{\mathrm{IV}}
		\end{bmatrix} \begin{bmatrix} 
			\tilde{X}\\
			\tilde{Y}\\
			\alpha_X\\
			\beta_Y
		\end{bmatrix} + \begin{bmatrix}
			\tilde{\boldsymbol{C}}_{\mathrm{I}} & \tilde{\boldsymbol{C}}_{\mathrm{II}}\\
			\tilde{\boldsymbol{C}}_{\mathrm{III}} & \tilde{\boldsymbol{C}}_{\mathrm{IV}}
		\end{bmatrix} \begin{bmatrix} 
			\tilde{\dot{X}}\\
			\tilde{\dot{Y}}\\
			\tilde{\dot{\alpha}}_X\\
			\tilde{\dot{\beta}}_Y
		\end{bmatrix} +  \begin{bmatrix}
			\tilde{\boldsymbol{M}}_{\mathrm{I}} & \tilde{\boldsymbol{M}}_{\mathrm{II}}\\
			\tilde{\boldsymbol{M}}_{\mathrm{III}} & \tilde{\boldsymbol{M}}_{\mathrm{IV}}
		\end{bmatrix} \begin{bmatrix} 
			\tilde{\ddot{X}}\\
			\tilde{\ddot{Y}}\\
			\tilde{\ddot{\alpha}}_X\\
			\tilde{\ddot{\beta}}_Y
		\end{bmatrix}.
	\end{split}
\end{gather}
\end{subequations}  
Within the equation, $\tilde{F}_X, \tilde{F}_Y$ and $\tilde{M}_X, \tilde{M}_Y$ are the induced forces and moments of the annulus acting on the rotor. The translational motion, velocity and acceleration of the rotor is given by $\tilde{X}, \tilde{\dot{X}}, \tilde{\ddot{X}}$ and $\tilde{Y},\tilde{\dot{Y}},\tilde{\ddot{Y}}$, whereas $\alpha_X, \tilde{\dot{\alpha}}_X, \tilde{\ddot{\alpha}}_X$ and $\beta_Y, \tilde{\dot{\beta}}_Y, \tilde{\ddot{\beta}}_Y$ denotes the angular motion, velocity and acceleration of the rotor around the $\tilde{X}$ and $\tilde{Y}$ axis. The describing coefficients depend on three different characteristics of the annulus: (i) the geometry of the annular gap, i.e. the gap length $\tilde{L}$, the shaft radius $\tilde{R}$, the mean gap clearance $\tilde{\bar{h}}$ and the gap function  $\tilde{h} = \tilde{h}\,(\varphi, \tilde{z}, \tilde{t})$ with the circumferential and axial coordinate $\varphi, \tilde{z}$; (ii) the operating conditions, i.e. the eccentric and angular position of the shaft $\tilde{e}, \alpha_X, \beta_Y$, the distance of the centre of rotation $\tilde{z}_T$ from the annulus entrance, the mean axial velocity $\tilde{\bar{C}}_z$, the angular velocity of the shaft $\tilde{\Omega}$ and the pre-swirl velocity at the annulus inlet $\tilde{C}_\varphi|_{z=0}$; (iii) the lubricant characteristics, i.e. the dynamic viscosity $\tilde{\eta}$ and the fluid density $\tilde{\varrho}$. On dimensional ground, the dimensionless rotordynamic coefficients are only a function of 8 dimensionless measures: (i) the dimensionless annulus length $L:=\tilde{L}/\tilde{R}$, (ii) the relative eccentricity $\varepsilon:=\tilde{e}/\tilde{\bar{h}}$, (iii, iv + v) the normalised angular displacements $\alpha:=\tilde{L} \, \alpha_X/\tilde{\bar{h}}$ and $\beta:=\tilde{L} \, \beta_Y/\tilde{\bar{h}}$ around the dimensionless fulcrum $z_T:=\tilde{z}_T/\tilde{L}$, (vi) the modified Reynolds number in circumferential direction $\psi Re_\varphi^*:= (\tilde{\bar{h}}/\tilde{R}) \,  (\tilde{\Omega} \tilde{R} \tilde{\bar{h}} / \tilde{\nu})^{n_\mathrm{f}}$, (vii) the flow number $\phi:=\tilde{\bar{C}}_z/(\tilde{\Omega}\tilde{R})$, (viii) the dimensionless pre-swirl $C_\varphi\vert_{z=0}:=\tilde{C}_\varphi\vert_{z=0}/(\tilde{\Omega}\tilde{R})$, cf. \cite{Kuhr.2022b}. Here, $\psi:=\tilde{\bar{h}}/\tilde{R}$ is the gap clearance and $n_\mathrm{f}$ is an empirical constant describing an arbitrary line within the double logarithmic Moody diagram 
\begin{equation}\label{eqn:dependenciesRotordynamicCoef}
	\begin{split}
		K_{ij}, C_{ij}&, M_{ij} = \mathrm{f}\left(L, \varepsilon, \alpha, \beta, z_T, Re^*_\varphi, \phi, C_\varphi\vert_{z=0}\right).
	\end{split}
\end{equation}

The dimensionless rotordynamic coefficients are separately defined for the forces and moments acting on the rotor due to translational and angular motions
\begin{equation}
		K_\mathrm{I} := \frac{2\tilde{\bar{h}}\tilde{K}_{ij}}{\tilde{\varrho}\tilde{\Omega}^2\tilde{R}^3\tilde{L}}, \quad K_\mathrm{II} := \frac{2\tilde{\bar{h}}\tilde{K}_\mathrm{II}}{\tilde{\varrho}\tilde{\Omega}^2\tilde{R}^3\tilde{L}^2}, \quad  K_\mathrm{III} := \frac{2\tilde{\bar{h}}\tilde{K}_\mathrm{III}}{\tilde{\varrho}\tilde{\Omega}^2\tilde{R}^3\tilde{L}^2}, \quad K_\mathrm{IV} := \frac{2\tilde{\bar{h}}\tilde{K}_\mathrm{IV}}{\tilde{\varrho}\tilde{\Omega}^2\tilde{R}^3\tilde{L}^3}.
\end{equation}

Here, the first two stiffness coefficients represent dynamic force coefficients due to translational and angular motion, whereas the later ones represent the moment coefficients due to translational and angular motion. The indices represent the corresponding sub-matrices of equation \ref{eqn:introduction_full_matrix}. The damping and inertia terms are defined accordingly
\begin{subequations}
	\begin{gather}
		C_\mathrm{I} := \frac{2\tilde{\bar{h}}\tilde{C}_\mathrm{I}}{\tilde{\varrho}\tilde{\Omega}\tilde{R}^3\tilde{L}}, \quad C_\mathrm{II} := \frac{2\tilde{\bar{h}}\tilde{C}_\mathrm{II}}{\tilde{\varrho}\tilde{\Omega}\tilde{R}^3\tilde{L}^2}, \quad C_\mathrm{III} := \frac{2\tilde{\bar{h}}\tilde{C}_\mathrm{III}}{\tilde{\varrho}\tilde{\Omega}\tilde{R}^3\tilde{L}^2}, \quad C_\mathrm{IV} := \frac{2\tilde{\bar{h}}\tilde{C}_\mathrm{IV}}{\tilde{\varrho}\tilde{\Omega}\tilde{R}^3\tilde{L}^3}, \\
		M_\mathrm{I} := \frac{2\tilde{\bar{h}}\tilde{M}_\mathrm{I}}{\tilde{\varrho}\tilde{R}^3\tilde{L}}, \quad M_\mathrm{II} := \frac{2\tilde{\bar{h}}\tilde{M}_\mathrm{II}}{\tilde{\varrho}\tilde{R}^3\tilde{L}^2}, \quad
		M_\mathrm{III} := \frac{2\tilde{\bar{h}}\tilde{M}_\mathrm{III}}{\tilde{\varrho}\tilde{R}^3\tilde{L}^2}, \quad M_\mathrm{IV} := \frac{2\tilde{\bar{h}}\tilde{M}_\mathrm{IV}}{\tilde{\varrho}\tilde{R}^3\tilde{L}^3}.
	\end{gather}
\end{subequations}

As stated before, the vast majority of experiments focus solely on the dynamic characteristics due to translational motion, i.e. sub-matrix I, of either journal bearings or annular seals, cf. \cite{Glienicke.1966,Childs.1982b,Nordmann.1984,Childs.1986c,Childs.1994,Marquette.1997}. Additionally, an adequate uncertainty quantification regarding the systematic and statistic uncertainty of the measurement signal is often missing when presenting the experimental data. For a detailed list of the theoretical and experimental investigations, see the review articles by \citet{Sternlicht.1967, Lund.1974, Pinkus.1987, Tiwari.2004, Tiwari.2005, Zakharov.2010}. 

\subsection{Experimental identification of all 48 coefficients}
In the following, we focus only on the experimental identification of all 48 rotordynamic coefficients, i.e. the dynamic characteristics including the tilt and moment coefficients. One of the first experimental identification of the all coefficients of an annular seal was presented by \citet{Kanemori.1992, Kanemori.1994b}. Here a test rig was presented, consisting of two rotors embedded in each other. By bearing and driving both rotors independently of each other, small translational and angular motions inducing frequency-dependent forces and moments on the rotor are imposed around the static equilibrium position By measuring both the motion of the rotor and the induced forces and moments on the rotor, the rotordynamic coefficients are identified using parameter identification methods, cf. \citep{Kanemori.1992, Tiwari.2018}. Essentially, the influence of the axial pressure difference $\Delta \tilde{p}$ as well as the angular frequency of the rotor $\tilde{\Omega}$ and the pre-swirl $\tilde{C}_\varphi |_{z=0}$ and the combination of cylindrical and conical whirls is investigated.

In contrast to the two embedded rotors used by Kanemori and Iwatsubo, \citet{Neumer.1994} and \citet{Matros.1995} use active magnetic bearings (AMBs) to position and excite the rotor. Here, the main advantage is that the bearings act as actuators and sensors at the same time. By measuring the electric current of the electromagnets as well as the rotor position, the force of the bearing is determined based on the force-current relation. This three-dimensional characteristic can either be determined by an extensive calibration procedure or by  calculation of the partial derivation of the field energy stored in the volume of the air gap with respect to the air gap itself, cf. \citet{Maslen.2009}. However, the force-current relation of the bearing is often linearised resulting in a significantly high uncertainty of the measured forces. Furthermore, it is only valid at the main coordinate axes of the magnetic bearing. Within their studies, Neumer and Matros et al. focus mainly on the influence of the angular frequency of the rotor $\tilde{\Omega}$ as well as the influence of the axial pressure difference $\Delta \tilde{p}$ on the rotordynamic coefficients. Here, in addition to the stiffness, damping and inertia coefficients from the induced forces due to translational motion, only stiffness and damping coefficients from the induced moments on the rotor due to translational motion are given. 

The aforementioned references are the exclusive bases for all 48 rotordynamic coefficients. Accordingly, there is a need for further reliable experimental investigations of the dynamic properties, especially regarding the additional rotordynamic tilt and moment coefficients and their dependencies according to equation \ref{eqn:dependenciesRotordynamicCoef}. For this purpose, a test rig originally presented by \citet{Kuhr.2022b} is extended in the following sections for the experimental determination of the dynamic force and moment characteristic using four linear independent whirling motions.

\section{Annular gap flow test rig}
\begin{figure}
	\centering
	\includegraphics[width=\textwidth]{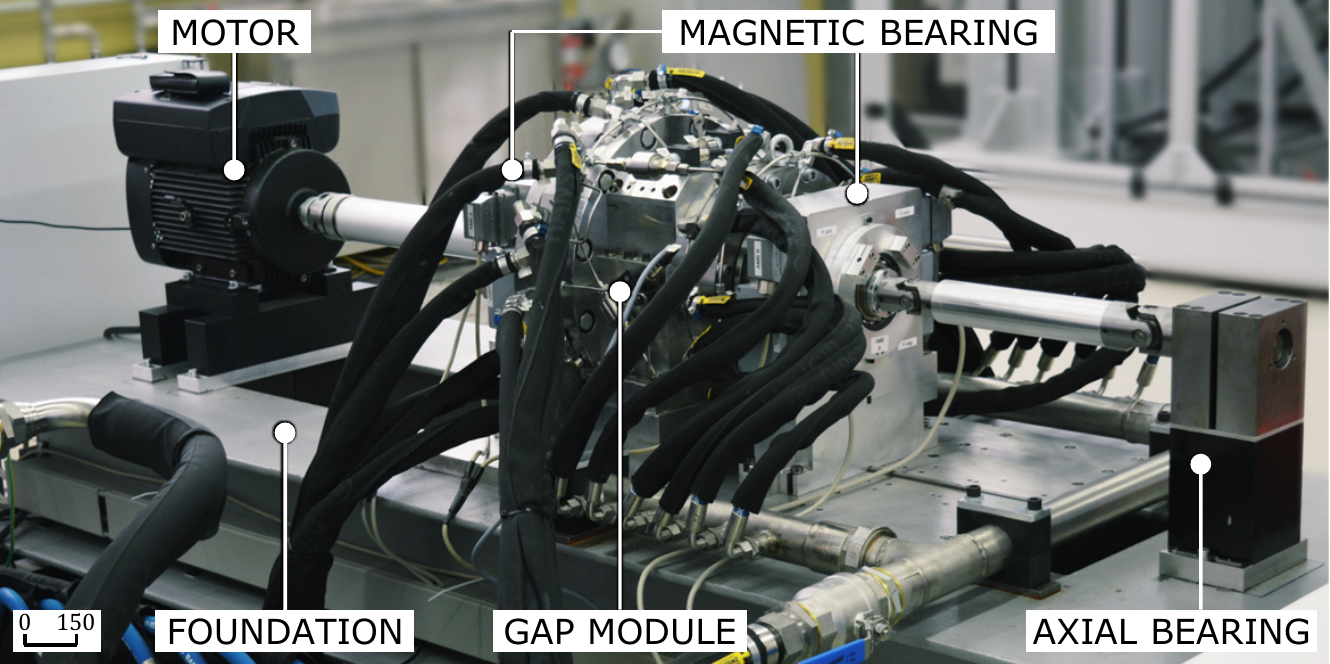}
	\caption{The annular gap test rig at the Chair of Fluid Systems at the Technische Universität Darmstadt, cf. \citet{Kuhr.2022, Kuhr.2022b}.}
	\label{fig:exp_test_rig}
\end{figure}

\cite{Kuhr.2022, Kuhr.2022b} present a test rig to experimentally investigate the static force characteristics of generic annuli, cf. figure \ref{fig:exp_test_rig}. The test rig mainly consists of two active magnetic bearings supporting the rotor and to act as an inherent displacement, excitation and force measurement system. In contrast to the test rigs used by Neumer and Matros et al., the force is obtained by measuring the magnetic flux density $\tilde{B}$ in the air gap between the rotor and the active magnetic bearing. Therefore, each pole of each electromagnet is equipped with a hall sensor. The force applied per pole is proportional to the magnetic flux density $\tilde{F}_{H} \propto \tilde{B}^2$. The test rig is designed to investigate annuli with relative lengths in a range of $0.2 \le L \le 1.8$ and relative clearances $10^{-3} \le \psi \le 10^{-2}$. Furthermore, it is capable of applying pressure differences across the annulus up to $13 \, \mathrm{bar}$, resulting in flow numbers up to $\phi \le 5$. For a detailed description of the test rig, its calibration and the measurement uncertainty of the used equipment, refer to \cite{Kuhr.2022, Kuhr.2022b}. In contrast to the formerly presented solely static force results, the test rig is extended to experimentally determine the dynamic force and moment characteristics of the investigated annuli. 

\subsection{Identification using four linear independent whirling motions}
Since the rotordynamic force and moment characteristics cannot be measured directly, they must be determined by using parameter identification methods, i.e. linear and quadratic regressions. The method used here is similar to that of \cite{Childs.1994} and requires the excitation of the rotor at user-defined frequencies and amplitudes. This is made possible by controlling the electric currents within the active magnetic bearings utilizing a feed-forward compensation (FFC) embedded in the controller. Subsequently, the induced dynamic forces and moments acting on the rotor $\tilde{F}_X$, $\tilde{F}_Y$ and $\tilde{M}_X$, $\tilde{M}_Y$ as well as the motion of the rotor in the four degrees of freedom, $\tilde{X}, \tilde{Y}$ and $\alpha_X, \beta_Y$ are measured. The measurement data acquisition of all sensors is carried out simultaneously based on a multifunction I\/O module with a sampling rate of $6000 \, \mathrm{Hz}$ per channel over a duration of five seconds. 

To identify the rotordynamic coefficients, the generalised equation of motion \ref{eqn:introduction_full_matrix} is first transferred into the frequency domain, yielding the complex equation of motion
\begin{equation}\label{eqn:complex_stiffness_method}
	\begin{split}
		-\begin{bmatrix}
			\tilde{\boldsymbol{\mathcal{F}}}_{X}\\
			\tilde{\boldsymbol{\mathcal{F}}}_{Y}\\
			\tilde{\boldsymbol{\mathcal{M}}}_{X}\\
			\tilde{\boldsymbol{\mathcal{M}}}_{Y}
		\end{bmatrix} = \begin{bmatrix}
			\tilde{\mathcal{K}}_{XX} &  \tilde{\mathcal{K}}_{XY} & \tilde{\mathcal{K}}_{X\alpha} & \tilde{\mathcal{K}}_{X \beta}\\
			\tilde{\mathcal{K}}_{XY} &  \tilde{\mathcal{K}}_{YY} & \tilde{\mathcal{K}}_{Y\alpha} & \tilde{\mathcal{K}}_{Y \beta}\\
			\tilde{\mathcal{K}}_{\alpha X} &  \tilde{\mathcal{K}}_{\alpha Y} & \tilde{\mathcal{K}}_{\alpha\alpha} & \tilde{\mathcal{K}}_{\alpha\beta}\\
			\tilde{\mathcal{K}}_{\beta X} &  \tilde{\mathcal{K}}_{\beta Y} & \tilde{\mathcal{K}}_{\beta\alpha} & \tilde{\mathcal{K}}_{\beta\beta}
		\end{bmatrix} \begin{bmatrix} 
			\tilde{\boldsymbol{\mathcal{D}}}_{X}\\
			\tilde{\boldsymbol{\mathcal{D}}}_{Y}\\
			{\boldsymbol{\mathcal{D}}}_{\alpha_X}\\
			{\boldsymbol{\mathcal{D}}}_{\beta_Y}
		\end{bmatrix}.
	\end{split}
\end{equation} 

Here, $\tilde{\mathcal{K}}_{ij}$ are the complex stiffness coefficients depending on the precessional frequency $\tilde{\omega}$. $\tilde{\mathcal{F}} = \mathfrak{F}(\tilde{F})$ and $\tilde{\mathcal{M}} = \mathfrak{F}(\tilde{M})$ are the Fourier transformations of the induces forces and moments on the rotor, whereas $\tilde{\mathcal{D}}_i = \mathfrak{F}(\tilde{i})$ are the Fourier transforms of the translational and angular excitation amplitudes. The real part of the complex stiffness $\Re \left({\tilde{\mathcal{K}}}_{ij}\right)$ contains the stiffness and inertia coefficient, whereas the imaginary part $ \Im\left(\tilde{{\mathcal{K}}}_{ij}\right)$ contains the damping coefficients
\begin{subequations}\label{eqn:complex_stiffness}
    \begin{gather}
        \tilde{\mathcal{K}}_{ij} = \tilde{K}_{ij} - \tilde{M}_{ij} \,  \tilde{\omega}^2 + \mathrm{i} \, \tilde{C}_{ij} \, \tilde{\omega}, \\
        \Re \left( \tilde{\mathcal{K}}_{ij} \right) = \tilde{K}_{ij} - \tilde{M}_{ij} \, \tilde{\omega}^2, \quad \Im \left( \tilde{\mathcal{K}}_{ij} \right) = \tilde{C}_{ij} \, \tilde{\omega}.
    \end{gather}
\end{subequations}

\begin{figure}
	\centering
	\includegraphics[scale=1.0]{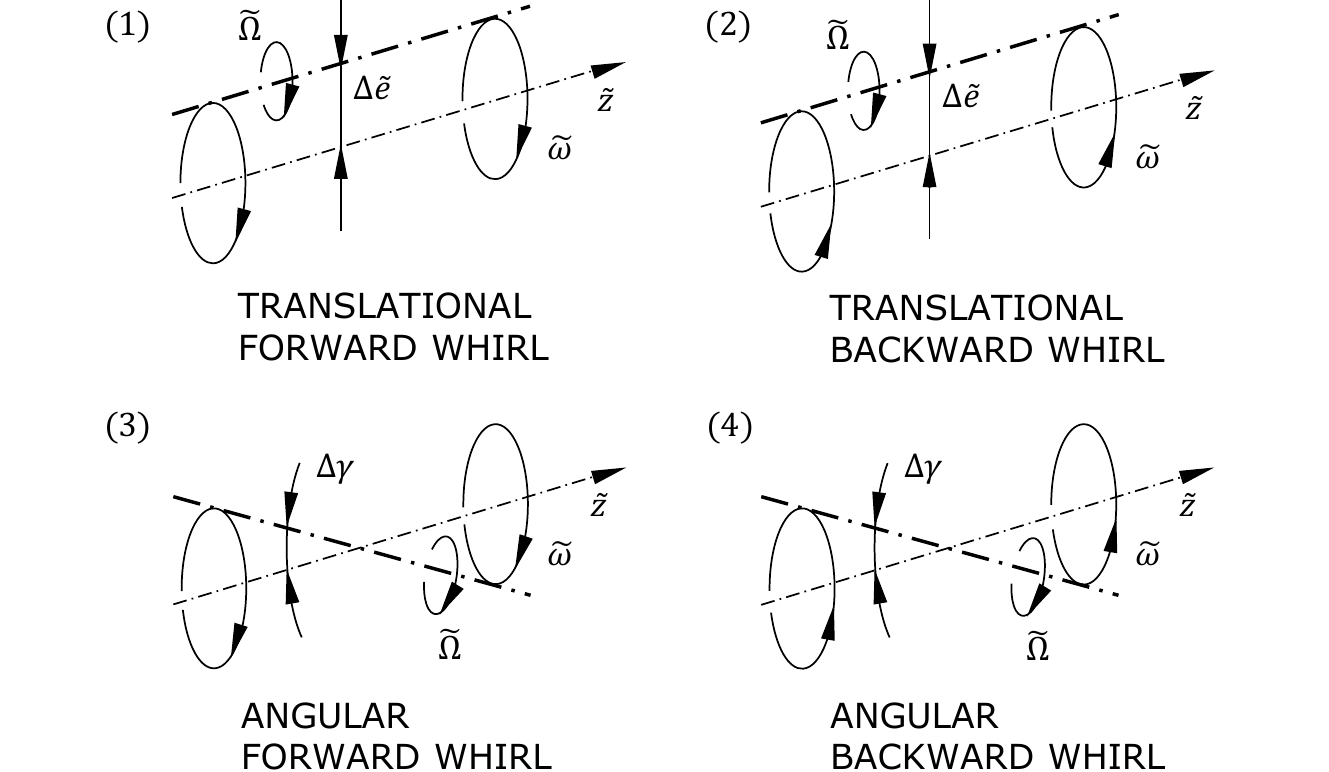}
	\caption{Linear independent whirling motions of the rotor: (1) translational excitation in the direction of rotation of the rotor, i.e. translational forward whirl; (2) translational excitation against the direction of rotation of the rotor, i.e. translational backward whirl; (3) angular excitation in the direction of rotation of the rotor, i.e. angular forward whirl; (4) angular excitation against the direction of rotation of the rotor, i.e. angular backward whirl.}
	\label{fig:linear_independent_excitations}
\end{figure}

It can easily be seen, that the stiffness and inertia coefficients can be obtained by applying a quadratic regression if the rotor is excited at different precessional frequencies. In contrast, a linear regression gives the damping of the annulus. The whole dynamic force and moment characteristic of the annulus, i.e. the 48 rotordynamic coefficients, can therefore be acquired by determining the 16 complex stiffnesses. However, to solve the underdetermined system of equations, four linear independent excitations of the rotor are needed. The linear excitation forms used here are shown in figure \ref{fig:linear_independent_excitations}. By using the active magnetic bearings, the rotor is forced into (1) a translational excitation in the direction of rotation of the rotor, i.e. a translational forward whirl, (2) a translational excitation against the direction of rotation of the rotor, i.e. a translational backward whirl, (3) an angular excitation in the direction of rotation of the rotor, i.e. an angular forward whirl, and (4) an angular excitation against the direction of rotation of the rotor, i.e. an angular backward whirl, with the user-defined amplitudes $\Delta \tilde{e}$ and $\Delta \gamma$ at the precessional frequency $\tilde{\omega} = 2\pi\tilde{f}$ around the static equilibrium position. 
Here, the angular excitation with amplitude $\Delta \gamma$ is established by applying a $180^\circ$ phase difference between the two active magnetic bearings. 

The four linear independent whirling motions yield to a total of 16 independent equations for the complex stiffnesses 
\begin{equation}\label{eqn:identification_full_matrix}
	\begin{split}
		-\begin{bmatrix}
			\tilde{\boldsymbol{\mathcal{F}}}_{X,1} & \tilde{\boldsymbol{\mathcal{F}}}_{X,2} & \tilde{\boldsymbol{\mathcal{F}}}_{X,3} & \tilde{\boldsymbol{\mathcal{F}}}_{X,4}\\
			\tilde{\boldsymbol{\mathcal{F}}}_{Y,1} & \tilde{\boldsymbol{\mathcal{F}}}_{Y,2} & \tilde{\boldsymbol{\mathcal{F}}}_{Y,3} & \tilde{\boldsymbol{\mathcal{F}}}_{Y,4}\\
			\tilde{\boldsymbol{\mathcal{M}}}_{X,1} & \tilde{\boldsymbol{\mathcal{M}}}_{X,2} & \tilde{\boldsymbol{\mathcal{M}}}_{X,3} & \tilde{\boldsymbol{\mathcal{M}}}_{X,4}\\
			\tilde{\boldsymbol{\mathcal{M}}}_{Y,1} & \tilde{\boldsymbol{\mathcal{M}}}_{Y,2} & \tilde{\boldsymbol{\mathcal{M}}}_{Y,3} & \tilde{\boldsymbol{\mathcal{M}}}_{Y,4}
		\end{bmatrix} = \begin{bmatrix}
			\tilde{\mathcal{K}}_{XX} &  \tilde{\mathcal{K}}_{XY} & \tilde{\mathcal{K}}_{X\alpha} & \tilde{\mathcal{K}}_{X \beta}\\
			\tilde{\mathcal{K}}_{XY} &  \tilde{\mathcal{K}}_{YY} & \tilde{\mathcal{K}}_{Y\alpha} & \tilde{\mathcal{K}}_{Y \beta}\\
			\tilde{\mathcal{K}}_{\alpha X} &  \tilde{\mathcal{K}}_{\alpha Y} & \tilde{\mathcal{K}}_{\alpha\alpha} & \tilde{\mathcal{K}}_{\alpha\beta}\\
			\tilde{\mathcal{K}}_{\beta X} &  \tilde{\mathcal{K}}_{\beta Y} & \tilde{\mathcal{K}}_{\beta\alpha} & \tilde{\mathcal{K}}_{\beta\beta}
		\end{bmatrix} \begin{bmatrix} 
			\tilde{\boldsymbol{\mathcal{D}}}_{X,1} & \tilde{\boldsymbol{\mathcal{D}}}_{X,2} & \tilde{\boldsymbol{\mathcal{D}}}_{X,3} & \tilde{\boldsymbol{\mathcal{D}}}_{X,4}\\
			\tilde{\boldsymbol{\mathcal{D}}}_{Y,1} & \tilde{\boldsymbol{\mathcal{D}}}_{Y,2} & \tilde{\boldsymbol{\mathcal{D}}}_{Y,3} & \tilde{\boldsymbol{\mathcal{D}}}_{Y,4}\\
			{\boldsymbol{\mathcal{D}}}_{\alpha_X,1} & {\boldsymbol{\mathcal{D}}}_{\alpha_X,2} & {\boldsymbol{\mathcal{D}}}_{\alpha_X,3} & {\boldsymbol{\mathcal{D}}}_{\alpha_X,4}\\
			{\boldsymbol{\mathcal{D}}}_{\beta_Y,1} & {\boldsymbol{\mathcal{D}}}_{\beta_Y,2} & {\boldsymbol{\mathcal{D}}}_{\beta_Y,3} & {\boldsymbol{\mathcal{D}}}_{\beta_Y,4}
		\end{bmatrix}.
	\end{split}
\end{equation}  
It should be noted that the measured variables of the forces and moments on the rotor as well as the measured variables of the motion of the rotor in the four degrees of freedom do not represent scalar variables after the Fourier transformation. Rather, the measured variables are arranged in a column matrix with $N$ entries for the positive and negative frequency components. However, the parameter identification is carried out solely on discrete data points at different precessional frequencies. Therefore, a linear regression is carried out to calculate the transfer function, i.e. the discrete points, of the complex stiffnesses before the actual parameter identification is carried out 
\begin{equation}\label{eqn:komplexe_Steifigkeiten_Skalar}
	\begin{split}
		& \begin{bmatrix}
			\tilde{\mathcal{K}}_{XX} &  \tilde{\mathcal{K}}_{XY} & \tilde{\mathcal{K}}_{X\alpha} & \tilde{\mathcal{K}}_{X \beta}\\
			\tilde{\mathcal{K}}_{XY} &  \tilde{\mathcal{K}}_{YY} & \tilde{\mathcal{K}}_{Y\alpha} & \tilde{\mathcal{K}}_{Y \beta}\\
			\tilde{\mathcal{K}}_{\alpha X} &  \tilde{\mathcal{K}}_{\alpha Y} & \tilde{\mathcal{K}}_{\alpha\alpha} & \tilde{\mathcal{K}}_{\alpha\beta}\\
			\tilde{\mathcal{K}}_{\beta X} &  \tilde{\mathcal{K}}_{\beta Y} & \tilde{\mathcal{K}}_{\beta\alpha} & \tilde{\mathcal{K}}_{\beta\beta}
		\end{bmatrix} = -\begin{bmatrix}
			\tilde{\boldsymbol{\mathcal{F}}}_{X,1} & \tilde{\boldsymbol{\mathcal{F}}}_{X,2} & \tilde{\boldsymbol{\mathcal{F}}}_{X,3} & \tilde{\boldsymbol{\mathcal{F}}}_{X,4}\\
			\tilde{\boldsymbol{\mathcal{F}}}_{Y,1} & \tilde{\boldsymbol{\mathcal{F}}}_{Y,2} & \tilde{\boldsymbol{\mathcal{F}}}_{Y,3} & \tilde{\boldsymbol{\mathcal{F}}}_{Y,4}\\
			\tilde{\boldsymbol{\mathcal{M}}}_{X,1} & \tilde{\boldsymbol{\mathcal{M}}}_{X,2} & \tilde{\boldsymbol{\mathcal{M}}}_{X,3} & \tilde{\boldsymbol{\mathcal{M}}}_{X,4}\\
			\tilde{\boldsymbol{\mathcal{M}}}_{Y,1} & \tilde{\boldsymbol{\mathcal{M}}}_{Y,2} & \tilde{\boldsymbol{\mathcal{M}}}_{Y,3} & \tilde{\boldsymbol{\mathcal{M}}}_{Y,4}
		\end{bmatrix}/ \begin{bmatrix} 
			\tilde{\boldsymbol{\mathcal{D}}}_{X,1} & \tilde{\boldsymbol{\mathcal{D}}}_{X,2} & \tilde{\boldsymbol{\mathcal{D}}}_{X,3} & \tilde{\boldsymbol{\mathcal{D}}}_{X,4}\\
			\tilde{\boldsymbol{\mathcal{D}}}_{Y,1} & \tilde{\boldsymbol{\mathcal{D}}}_{Y,2} & \tilde{\boldsymbol{\mathcal{D}}}_{Y,3} & \tilde{\boldsymbol{\mathcal{D}}}_{Y,4}\\
			{\boldsymbol{\mathcal{D}}}_{\alpha_X,1} & {\boldsymbol{\mathcal{D}}}_{\alpha_X,2} & {\boldsymbol{\mathcal{D}}}_{\alpha_X,3} & {\boldsymbol{\mathcal{D}}}_{\alpha_X,4}\\
			{\boldsymbol{\mathcal{D}}}_{\beta_Y,1} & {\boldsymbol{\mathcal{D}}}_{\beta_Y,2} & {\boldsymbol{\mathcal{D}}}_{\beta_Y,3} & {\boldsymbol{\mathcal{D}}}_{\beta_Y,4}
		\end{bmatrix}.
	\end{split}
\end{equation}
Subsequently, the rotordynamic coefficients are identified via a second regression, i.e. a quadratic one for the stiffness and inertia coefficients and a linear one for the damping coefficients of the annuls. The regression is based on an adapted version of the regression procedure by \citet{Jolly.2018}. Here, the regression is extended by an iterative procedure to eliminate possible measurement outliers based on a robust bisquare weighting, cf. \citet{TheMathWorksInc..2021}. The iteratively determined weights $W$ map the contribution of each measurement point to the regression curve
\begin{subequations}\label{eqn:bisquare}
    \begin{gather}
        \begin{split}
            \begin{bmatrix}
            \sum \limits_{k = 1}^{m} W_k & -\sum \limits_{k = 1}^{m} W_k \, \tilde{\omega}_{k}^{2} \\
            -\sum \limits_{k = 1}^{m} W_k \, \tilde{\omega}_{k}^{2} & \sum \limits_{k = 1}^{m} W_k \, \tilde{\omega}_{k}^{4}
        \end{bmatrix} 
        \begin{bmatrix}
            \tilde{K}_{ij} \\
            \tilde{M}_{ij}
        \end{bmatrix} = 
        \begin{bmatrix}
            \sum \limits_{k = 1}^{m} W_k \, \Re\left[ \tilde{\mathcal{K}}_{ij} \left( \tilde{\omega}_{k} \right) \right] \\
            -\sum \limits_{k = 1}^{m} W_k \, \tilde{\omega}_{k}^{2} \, \Re \left[ \tilde{\mathcal{K}}_{ij} \left( \tilde{\omega}_{k} \right) \right]
        \end{bmatrix}, 
        \end{split} \\
        \tilde{C}_{ij} = \frac{\sum \limits_{k = 1}^{m} W_k \, \tilde{\omega}_{k} \, \Im \left[ \tilde{\mathcal{K}}_{ij} \left( \tilde{\omega}_{k} \right) \right]}{\sum \limits_{k = 1}^{m} W_k \, \tilde{\omega}_{k}^{2}}.
    \end{gather}
\end{subequations}
Within the regression procedure, the index $k$ represents the number of individual precessional frequencies $\tilde{\omega}_k$. If a value, i.e. a discrete complex stiffness at a particular precessional frequency, lies significantly outside the regression curve, its weight is reduced in comparison to the other supposedly correct values, thus reducing its influence on the regression curve. The procedure is repeated until the change in weights falls below a previously defined value. Here, a change of less than 1 permil per iteration in the weights is used as convergence criterion. 

To solve the system of equations~\ref{eqn:bisquare} at least two different angular frequencies $\tilde{\omega}_k$ are necessary. To improve the accuracy of the regression, at least seven angular frequencies are used to identify the rotordynamic coefficients presented in section~\ref{sec:results}. With the independent motions necessary for identification, this results in a number of $28$ individual measurements for the identification of all $48$ rotordynamic coefficients.

\subsection{Uncertainty quantification}
The experimental identification of the rotordynamic coefficients is only suitable for validation purposes if the associated measurement uncertainty is also given in addition to the identified coefficients. However, many experimental investigations, especially for all 48 rotordynamic coefficients, are presented without an adequate uncertainty quantification, cf. \citet{Kanemori.1992, Neumer.1994, Matros.1995}, or take only the standard error of the y-data about the curve fit into account, cf. \citet{Moreland.2018}. For this reason, the assumptions used within the paper are subsequently presented. The evaluation is mainly based on the Guide to the expression of Uncertainty in Measurement (GUM) of the \citet{JointCommitteeforGuidesinMetrology.2008}.

\subsubsection{Systematic and statistical measurement uncertainty for oscilating signals}
For time invariant measurements, the empirical mean $\tilde{\bar{x}}$ of a measured data set is usually given in the form
\begin{equation}
    \tilde x = \tilde{\bar{x}} \pm \delta \tilde{x}\quad \mathrm{with} \quad \tilde{\bar{x}} = \frac{1}{N} \sum \limits_{i = 1}^N \tilde{x}_i.
\end{equation}
The measurement uncertainty of a particular data set $\delta ( \tilde{x} )$, e.g. the uncertainty of the rotor position $\delta ( \tilde{X} )$, is always composed of a systematic $\delta ( \tilde{x} )_{\mathrm{SYS}}$ and a statistical component $\delta ( \tilde{x} )_{\mathrm{STAT}}$. The systematic measurement uncertainty $\delta ( \tilde{x} )_{\mathrm{SYS}}$ results in a repeatable, uniform influence of the values contained in the data set. A repeated measurement of the physical quantity $\tilde{x}$ does not lead to a reduction of the systematic measurement uncertainty~$\delta ( \tilde{x} )_{\mathrm{SYS}}$. In contrast, the statistical measurement uncertainty $\delta ( \tilde{x} )_{\mathrm{STAT}}$ originates from random fluctuations of the measurement signal. The statistical measurement uncertainty~$\delta ( \tilde{x} )_{\mathrm{STAT}}$ can therefore be reduced by repeating the measurement multiple times.

The systematic measurement uncertainty $\delta ( \tilde{x} )_{\mathrm{SYS}}$ of a measurement signal is generally determined from the associated data sheet of the sensor used. Here, the uncertainty of the sensor is given as the maximum deviation of the linearity of the sensor $\delta ( \tilde{x} )_{\mathrm{MAX}}$ from the actual sensor characteristic curve. This scale uncertainty represents a maximum measurement uncertainty of the sensor. The probability distribution of this maximum deviation is unknown in most cases. For this reason, the GUM always suggests the assumption of a uniform distribution for the treatment of the systematic measurement uncertainties $\delta ( \tilde{x} )_{\mathrm{SYS}}$ from data sheets. This can be converted to a normal distribution based on a standard deviation $\sigma ( \tilde{x} )_{\mathrm{SYS}}$ for the application of Gaussian error propagation.
\begin{equation}
    \sigma \left( \tilde{x} \right)_{\mathrm{SYS}} = \frac{\delta ( \tilde{x} )_{\mathrm{max}}}{ \sqrt{3} }.
\end{equation}
The systematic measurement uncertainty $\delta ( \tilde{x} )_{\mathrm{SYS}}$ is obtained by multiplying the standard deviation $\sigma ( \tilde{x} )_{\mathrm{SYS}}$ by the Student's $t$ factor
\begin{equation}
    \delta \left( \tilde{x} \right)_{\mathrm{SYS}} = t_{\mathrm{VI}} \, \sigma \left( \tilde{x} \right)_{\mathrm{SYS}}.
\end{equation}
Here, the $t$-factor $t_{\mathrm{VI}}$ results from the number of measured values in the data set $N$ as well as the specification of a confidence interval (VI). The use of the $t$ factor considers that through a random measurement of a physical quantity $\tilde{x}$, strictly speaking, normal distribution only exists for an infinitely long data set, i.e. $N \to \infty$. Finite measurements of the physical quantity result in a $t$-distributed data set. This is considered, by multiplying the standard deviation $\sigma ( \tilde{x} )_{\mathrm{SYS}}$ by the Student's $t$ factor $t_\mathrm{VI}$. Here, a two-sided confidence interval of $95 \, \%$ is assumed for the measurement uncertainty. The systematic measurement uncertainty~$\delta ( \tilde{x} )_{\mathrm{SYS}}$ is given by
\begin{equation}
    \delta \left( \tilde{x} \right)_{\mathrm{SYS}} = t_{95\%} \, \sigma \left( \tilde{x} \right)_{\mathrm{SYS}} = 1.96 \, \sigma \left( \tilde{x} \right)_{\mathrm{SYS}}.
\end{equation}

\begin{figure}
	\centering
	\includegraphics[width=13.5cm]{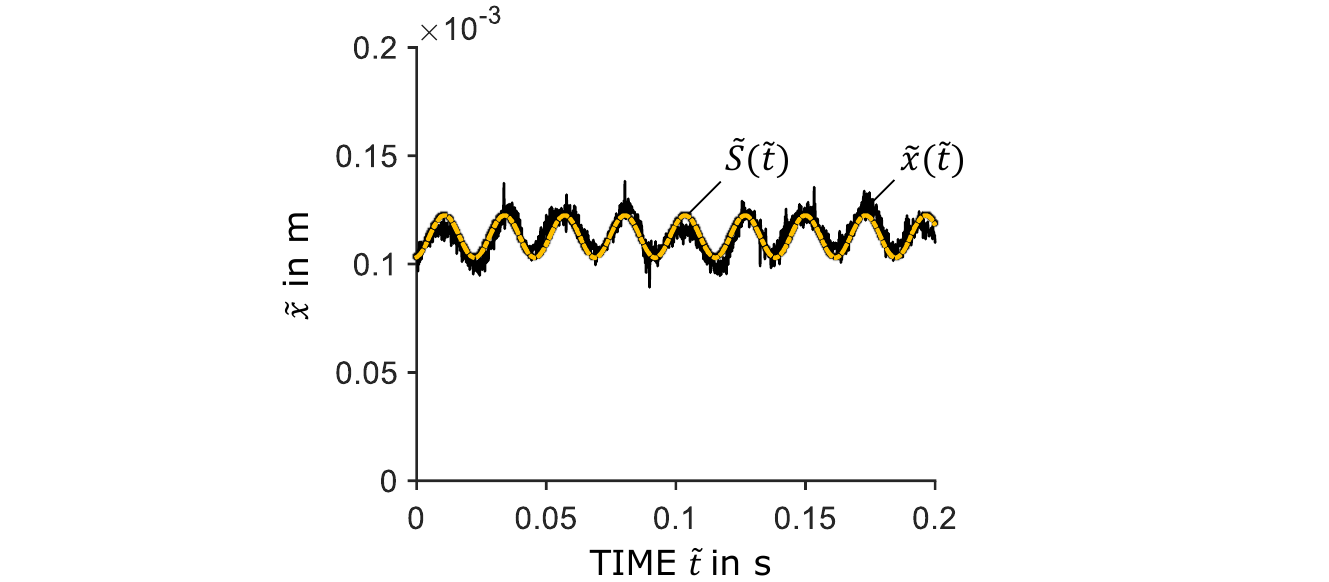}
	\caption{Approximation of the time-dependent measure $\tilde{x}$ using a harmonic function $\tilde{S}$.}
	\label{fig:figure_statistic_uncertainty}
\end{figure}

The statistical measurement uncertainty $\delta ( \tilde{x} )_{\mathrm{STAT}}$ of the measured data set ${\tilde{x}_i}$ is determined using the standard deviation of the data set $\sigma ( \tilde{x} )$, the Student's $t$ factor $t_{95\%}$ and the number of measured values in the data set $N$
\begin{equation}
    \delta \left( \tilde{x} \right)_{\mathrm{STAT}} = \frac{t_{95\%}}{\sqrt{N}} \sigma \left( \tilde{x} \right).
\end{equation}
The standard deviation of the data set of a stationary measure $ \delta \left( \tilde{x} \right)_{\mathrm{STAT}}$ is calculated using the empirical mean of the data set $\tilde{\bar{x}}$ to be
\begin{equation}
    \sigma \left( \tilde{x} \right) = \sqrt{ \frac{1}{N - 1} \sum \limits_{i = 1}^N \left( \tilde{x}_i - \tilde{\bar{x}} \right)^2 }.
\end{equation}
For transient signal, e.g. when exciting the rotor, the standard deviation $\sigma ( \tilde{x} )$ is significantly overestimated when by using the empirical mean $\tilde{\bar{x}}$. To overcome this, the standard deviation $\sigma ( \tilde{x} )$ in the context of the paper is calculated differently. Here, each time-dependent measurement signal is not approximated by the empirical mean $\tilde{\bar{x}}$, but by a harmonic function $\tilde{S}$ with defined amplitude $\tilde{a}$, angular frequency of the excitation $\tilde{\omega}$, phase shift $b$ and offset $\tilde{c}$, cf. figure~\ref{fig:figure_statistic_uncertainty}. 

\begin{equation}
    \tilde{S} = \tilde{a} \, \sin \left( \tilde{\omega} \tilde{t} + b \right) + \tilde{c}.
\end{equation}
 The standard deviation of the data set of a transient measure $\sigma ( \tilde{x} )$ is formed in analogy to the standard deviation of the steady-state signal
\begin{equation}
    \sigma \left( \tilde{x} \right) = \sqrt{ \frac{1}{N - 1} \sum \limits_{i = 1}^N \left( \tilde{x}_i - \tilde{S}_i \right)^2 }.
\end{equation}
Therefore, the main definition of the standard deviation $\sigma ( \tilde{x} )$, i.e the random dispersion around a signal, remains valid even for the harmonic measurement signals.

\subsubsection{The frequency domain}\label{subsec:freqeuncyDomain}
To this point, all aforementioned assumptions were made for signals within the time domain. However, the identification of the rotordynamic coefficients is carried out based on Fourier-transformed time signals, i.e in the frequency domain. For transferring the measurement uncertainty in the frequency domain, it is assumed that the total measurement uncertainty of the signal $\delta ( \tilde{x} ) =  \sqrt{\delta ( \tilde{x}_{\mathrm{SYS}} ) ^2 + \delta ( \tilde{x}_{\mathrm{STAT}} ) ^2}$ can be assigned to each discrete value of a time signal. Regarding the statistical measurement uncertainty~$\delta ( \tilde{x} )_{\mathrm{STAT}}$, this represents a conservative estimation of the overall measurement uncertainty. Furthermore, it is assumed that the overall measurement uncertainty can be regarded as an uncertainty in the scaling, i.e. the slope of the characteristic. Accordingly, when transforming the signal into the frequency domain, the uncertainty can be equally applied to each frequency of the signal. This results in a conservative estimation and constant measurement uncertainty $\delta [\mathcal{F} ( \tilde{x} ) ] = \mathrm{const}$ of the transformed signal $\mathcal{F} ( \tilde{x} )$ at all frequencies. If, on the other hand, the measurement uncertainty were not interpreted as a scale error but as a constant offset, the Fourier transform of the measurement signal would, with negligible statistical measurement uncertainty, only have an uncertainty at vanishing excitation frequencies, i.e. $\tilde{\omega} = 0$. Accordingly, the measurement uncertainty of the rotordynamic coefficients is calculated from the systematic and statistical measurement uncertainty of the measurement signals with subsequent Fourier transformation and the following Gaussian uncertainty propagation based on the equations of the identification procedure, cf. equation \ref{eqn:bisquare}. Conveniently, the calculation of the measurement uncertainty in the frequency domain is carried out using the open-source Matlab library METAS UncLib \citep{Zeier.2012}, provided by the Swiss Federal Institute of Metrology (METAS).

\newpage
\section{Preliminary testing}
Prior to the experimental investigations regarding the dynamic characteristics of the annulus, extensive preliminary tests are conducted to address the validity of the identification procedure as well as the reproducibility of the results. Furthermore, the dynamic influence of the test rig without an annulus has to characterised. The different angular frequencies for the excitation of the rotor $\tilde{\omega}$ during the preliminary test as well as the experimental investigations are chosen that they do not coincide with the angular frequency of the rotor $\tilde{\Omega}$ and have a sufficient distance to the natural frequency of the foundation and the rotor. While the natural frequency of the foundation is at $6 \, \mathrm{Hz}$, the rotor has its first bending mode at $910 \, \mathrm{Hz}$. To ensure a sufficient distance of the excitation frequencies from the natural frequencies, the excitation frequencies range from $\tilde{f} = \tilde{\omega}/(2\pi) = 10 \, \mathrm{Hz}$ to $50 \, \mathrm{Hz}$. The excitation amplitudes for the translational and angular excitation are set to $\Delta \tilde{X} = \Delta \tilde{Y} = 10 \times 10^{-6} \, \mathrm{m}$ and $\Delta \alpha_X = \Delta \beta_Y = 5.9 \times 10^{-3}\,^\circ$.

\subsection{Validation of the identification procedure}
To ensure a reliable and accurate determination of the rotordynamic coefficients, the presented identification procedure is validated using known quantities. Here, the known weighted mass of the rotor $\tilde{M}_{\mathrm{ROT}}$ is chosen as the first comparative quantity. For this purpose, only the rotor of the gap flow test rig is subjected to the identification procedure. 

The rotor mass is subsequently identified by applying the identification procedure. The rotor mass $\tilde{M}_{\mathrm{ROT}}$ is obtained by focusing on the real part of the complex stiffness $\Re ( \tilde{\mathcal{K}}_{YY})$. Since only the rigid rotor is subjected to identification, the stiffness and damping is negligible. 

\begin{figure}
	\centering
	\includegraphics[scale=1.0]{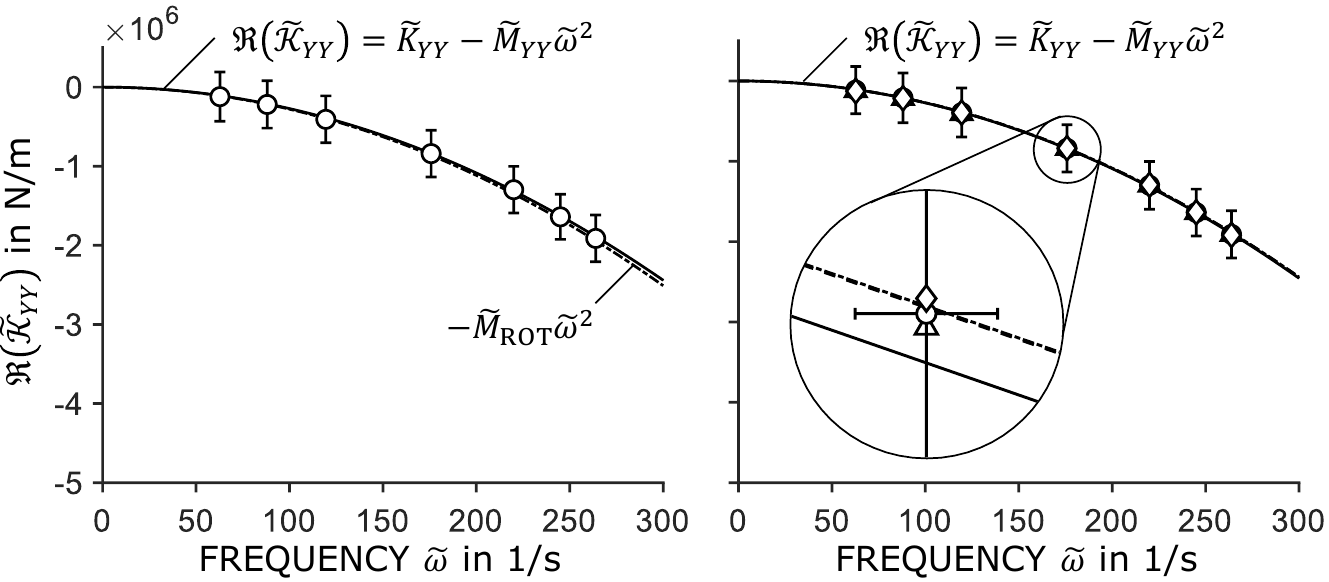}
	\caption{Identification of the rotor mass (left) and reproducibility of the identification on three different measurement days (right).}
	\label{fig:complex_stiffness_repeatability}
\end{figure}

The left side of Figure~\ref{fig:complex_stiffness_repeatability} exemplary shows the identified real part of the complex stiffness $\Re ( \tilde{\mathcal{K}}_{YY} ) $ as well as the calculated complex stiffness from the weighted rotor mass, i.e. $-\tilde{M}_{\mathrm{ROT}}\,\tilde{\omega}^2 $. The markers represent the applied precessional frequencies $\tilde{\omega}$, whereas the solid and dashed line represents the quadratic regression and the calculated complex stiffness of the weighted rotor mass. It can be seen that the regression is in excellent agreement with the calculated curve. Accordingly, only minor differences between the identified mass of the rotor $\tilde{M}_{YY}$ and the weighted mass of the rotor $\tilde{M}_{\mathrm{ROT}}$ occur. 

Table \ref{tab:identified_rotor_mass_and_inertia} lists the identified rotor mass $\tilde{M}_{XX} = \tilde{M}_{YY}$ and the identified rotor inertia $\tilde{M}_{\alpha \alpha} = \tilde{M}_{\beta \beta}$ compared to the weighted rotor mass~$\tilde{M}_{\mathrm{ROT}}$ and the rotor inertia~$\tilde{\Theta}_{\mathrm{ROT}}$ determined from a CAD model. It is shown that the rotor mass can be identified with a deviation of~$2.6\,\%$, whereas the inertia is determined with a deviation of $3.4\,\%$. 
At first, the given uncertainties within the identified coefficients seem rather high compared to the value of the coefficient itself. However, this is mainly due to the adequate uncertainty quantification, including both systematic and statistic measurement uncertainty not only in the time domain but also in the frequency domain, cf. section \ref{subsec:freqeuncyDomain}. In contrast, if only the statistic uncertainty is considered, the measurement uncertainty of the identified rotor mass reduces to $\delta(\tilde{M}_{XX}) = \delta(\tilde{M}_{YY}) = \pm 0.57 \, \mathrm{kg}$.
\begin{table}%
	\centering
	\caption{Identified rotor mass and inertia $\tilde{M}_{YY}$ and $\tilde{M}_{\alpha\alpha}$ in comparison with the weighted mass $\tilde{M}_\mathrm{ROT}$ and the inertia of the rotor determined from a CAD model $\tilde{\Theta}_\mathrm{ROT}$.}
	\label{tab:identified_rotor_mass_and_inertia}
	\small
    \begin{tabular}{p{2.8cm}|p{2.8cm}|p{3.3cm}|p{1.9cm}}
			$\tilde{M}_{XX} = \tilde{M}_{YY}$ & $\tilde{M}_{\mathrm{ROT}}$ & $\tilde{M}_{\alpha \alpha} = \tilde{M}_{\beta \beta}$ & $\tilde{\Theta}_{\mathrm{ROT}}$\\
			\toprule
			$\left( 27.16 \pm 4.67 \right) \, \mathrm{kg}$ & $\left( 27.89 \pm 0.06 \right) \, \mathrm{kg}$ & $\left( 0.900 \pm 0.392 \right) \, \mathrm{kg \, m^2}$ & $0.932 \, \mathrm{kg \, m^2}$\\
    \end{tabular}
\end{table}

In addition to the identification of the rotor mass $\tilde{M}_\mathrm{ROT}$, figure~\ref{fig:complex_stiffness_repeatability} on the right shows the reproducibility of the identification procedure on three different measurement days. As with the comparison between the identified and weighed rotor mass $\tilde{M}_{YY}$ and $\tilde{M}_\mathrm{ROT}$, the differences between the individual measurement days are hardly noticeable. The identification procedure thus represents a valid and reproducible method for the identification of the rotordynamic coefficients and can be applied for the identification of the dynamic properties of annular gaps.
However, Before the actual parameter studies are carried out, the dynamic influence of the test rig without an annulus has to be characterised. This is of particular importance since the actual measurements always consists of the dynamic characteristics of the test rig and the annulus, cf. \citet{Childs.1994}. Therefore, the dynamic influence of the test rig has to be known and subtracted for the test results to get the force and moment characteristics of the annuls.

\subsection{Dynamic characteristics of the water filled test rig without an annulus}\label{sec:characteristics_test_rig}
\begin{figure}
	\centering
	\includegraphics[scale=1.0]{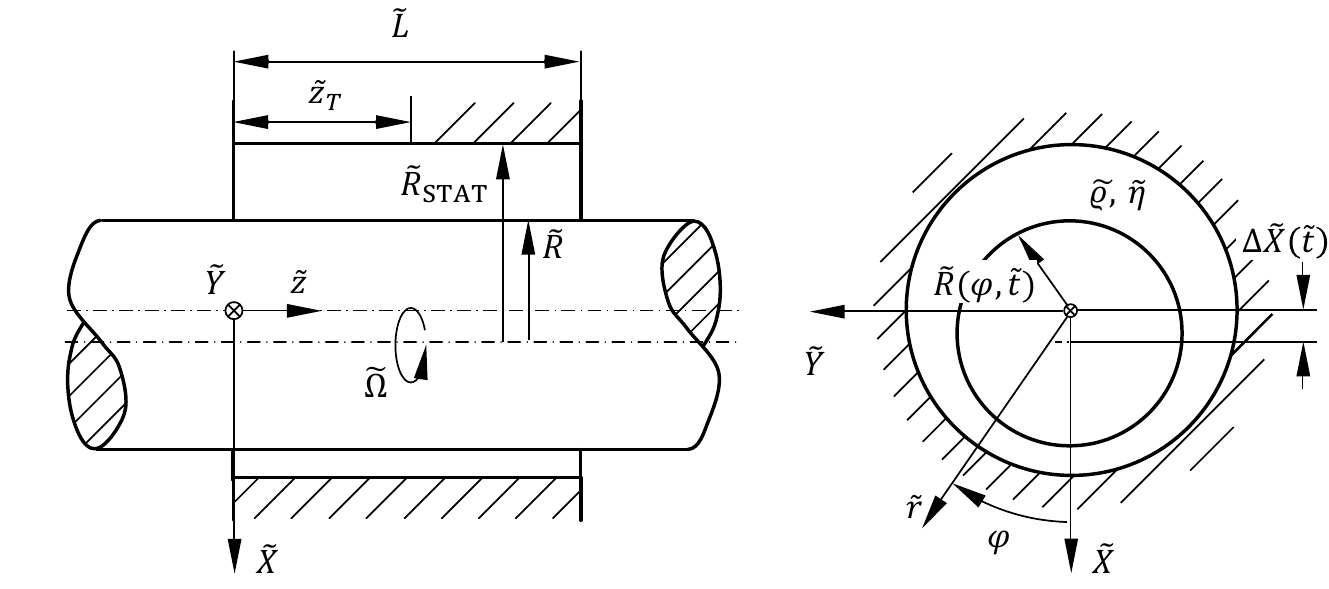}
	\caption{Oscillating and rotating rotor in an annulus filled with an inviscid fluid.}
	\label{fig:added_mass_and_lift_potential_flow}
\end{figure}
To characterise the dynamic properties of the experimental setup, the gap flow test rig is filled with water and the dynamic properties of the setup without an annulus are identified. Primarily, the added mass of the test rig $\tilde{M}_{\mathrm{VIRT}}$ is to be determined. Additionally, the influence of further parameters, such as the angular frequency of the rotor $\tilde{\Omega}$, the excitation amplitudes $\Delta \tilde{X} = \Delta \tilde{Y}$ and $\Delta \alpha_X = \Delta \beta_Y$ and the pressure within the test rig $\tilde{p}$, on the dynamic properties is investigated. The added mass can be determined both analytically, i.e. by using irrotational flow theory, cf. \citet{Spurk.1995,Batchelor.2012}, and experimentally. Therefore, besides the rotor mass, a further comparative quantity is available for the validation of the identification procedure.
For the analytical determination of the added mass $\tilde{M}_{\mathrm{VIRT}}$, the potential of the flow $\tilde{\Phi}$ of an oscillating and rotating rotor has to be determined, cf. figure~\ref{fig:added_mass_and_lift_potential_flow}. Here, the relative gap clearance is much larger than the one used in the annulus, i.e. $\psi \ll 1$. Moreover, the problem is treated as a plane flow, which leads to the assumption of a much larger length of the stator $\tilde{L}$ than its radius $R_\mathrm{STAT}$, i.e. $\tilde{L} \gg \tilde{R}_\mathrm{STAT}$. The potential of the flow in the annular gap $\tilde{\Phi}$ yields
\begin{equation}\label{eqn:potential_flow_annular_gap}
    \tilde{\Phi} = - \delta \tilde{\dot{X}} \frac{ \tilde{R}^{2} }{ \tilde{R}_{ \mathrm{STAT}}^{2} - \tilde{R}^{2} } \left( \tilde{r} + \frac{ \tilde{R}_{\mathrm{STAT}}^{2} }{ \tilde{r} } \right) \cos \varphi + \frac{ \tilde{\Gamma} }{ 2\pi } \varphi, \quad \mathrm{for} \quad \tilde{L} \gg \tilde{R}_{\mathrm{STAT}}.
\end{equation}

Here, $\tilde{\Gamma} \varphi / ( 2\pi )$ denotes the potential of a vortex filament with circulation $\tilde{\Gamma}$ along the $\tilde{z}$ axis. The radial and tangential velocities of the flow $\tilde{c}_r$ and $\tilde{c}_\varphi$ result from the respective derivatives of the potential
\begin{equation}
    \tilde{c}_{r} = \frac{\partial \tilde{\Phi}}{\partial \tilde{r}}=-\Delta\tilde{\dot{X}} \frac{\tilde{R}^{2}}{\tilde{R}_\mathrm{STAT}^{2}-\tilde{R}^{2}}\left(1-\frac{\tilde{R}_\mathrm{STAT}^{2}}{\tilde{r}^2}\right) \cos \varphi,
\end{equation}
\begin{equation}
    \tilde{c}_\varphi = \frac{1}{\tilde{r}}\frac{\partial \tilde{\Phi}}{\partial \varphi}=\delta\tilde{\dot{X}} \frac{\tilde{R}^{2}}{\tilde{R}_\mathrm{STAT}^{2}-\tilde{R}^{2}}\left(1+\frac{\tilde{R}_\mathrm{STAT}^{2}}{\tilde{r}^2}\right) \sin \varphi + \frac{\tilde{\Gamma}}{2\pi}\frac{1}{\tilde{r}}.
\end{equation}
The added mass $\tilde{M}_{\mathrm{VIRT}}$ can be calculated from the kinetic energy of the flow $\tilde{E}_\mathrm{KIN}$. The kinetic energy $\tilde{E}_{\mathrm{KIN}}$ as well as the resulting added mass $\tilde{M}_{\mathrm{VIRT}}$ is obtained with the volume of the rotor $\tilde{V} = \pi \tilde{L} \tilde{R}^2$
\begin{equation}
    \tilde{E}_\mathrm{KIN} = \tilde{R} \frac{\tilde{\varrho}}{2}\int\limits_{0}^{\tilde{L}} \int\limits_{0}^{2\pi} \tilde{\Phi} \frac{\partial \tilde{\Phi}}{\partial \tilde{r}} \, \mathrm{d} \varphi \, \mathrm{d} \tilde{z} = \pi \tilde{R}^{2}\tilde{L} \, \Delta \tilde{\dot{X}}^{2} \frac{\tilde{\varrho}}{2} \frac{\tilde{R}_\mathrm{STAT}^{2} + \tilde{R}^{2}}{\tilde{R}_\mathrm{STAT}^{2} - \tilde{R}^{2}}
\end{equation}
\begin{equation}
    \tilde{M}_{\mathrm{VIRT}} = \frac{2 \tilde{E}_\mathrm{KIN}}{\Delta \tilde{\dot{X}}^{2}} = \tilde{\varrho} \tilde{V} \frac{\tilde{R}_\mathrm{STAT}^{2} + \tilde{R}^{2}}{\tilde{R}_\mathrm{STAT}^{2} - \tilde{R}^{2}}.
\end{equation}
In addition to the influence of the added mass of water $\tilde{M}_{\mathrm{VIRT}}$, the rotor experiences a force $\tilde{F}_{\mathrm{LIFT}}$ perpendicular to its excitation. This lifting force is known from aerodynamics and is calculated from the integral of the dynamic pressure $\tilde{p}_{\mathrm{DYN}}$ on the rotor surface
\begin{equation}
    \tilde{F}_\mathrm{LIFT} = \tilde{L} \int \limits_{0}^{2 \pi} \tilde{p}_\mathrm{DYN} \tilde{R} \sin \varphi \, \mathrm{d} \varphi = \tilde{L} \tilde{R} \frac{\tilde{\varrho}}{2} \int \limits_{0}^{2 \pi} \left( \tilde{c}_{r}^{2} \big|_{r = R} + \tilde{c}_{\varphi}^{2} \big|_{r = R} \right) \sin \varphi \, \mathrm{d} \varphi.
\end{equation}
Due to the symmetry of the rotor, the radial component of the velocity $\tilde{c}_r$ vanishes on integration and the lift $\tilde{F}_\mathrm{LIFT}$ acting on the rotor is given by
\begin{equation}
    \tilde{F}_\mathrm{LIFT} = \tilde{L} \frac{\tilde{\varrho}}{2} \Delta \tilde{\dot{X}} \frac{\tilde{R}_\mathrm{STAT}^{2} + \tilde{R}^{2}}{\tilde{R}_\mathrm{STAT}^{2} - \tilde{R}^{2}} \tilde{\Gamma}.
\end{equation}

Comparing the added mass of water $\tilde{M}_{\mathrm{VIRT}}$ with the physical mass of water in the annular gap $\pi \tilde{\varrho} \tilde{L} (\tilde{R}_{\mathrm{STAT}}^2 - \tilde{R}^2 )$ it becomes clear that the added mass is not only the physical mass of the water in the annulus. Rather, a transmission ratio resulting from the descriptive radii $\tilde{R}$ and $\tilde{R}_\mathrm{STAT}$ is applied.

According to the Kutta-Joukowski theorem, cf. \citet{Spurk.2008, Batchelor.2012}, the lift is linearly proportional to the circulation $\tilde{\Gamma}$ as well as the velocity of the rotor due to its excitation $\Delta \tilde{\dot{X}}$. The circulation of the rotor is given by $\tilde{\Gamma} = 2\pi \tilde{\Omega} \tilde{R}^2$. In terms of the dynamic properties of axial flow annular gaps, the lift $\tilde{F}_\mathrm{LIFT}$ can be interpreted as a velocity-proportional force on the rotor perpendicular to the direction of motion. Accordingly, the lift $\tilde{F}_\mathrm{LIFT}$ related to the inflow velocity $\Delta \tilde{\dot{X}}$ can be regarded as cross-coupled damping coefficient $\tilde{C}_{\mathrm{LIFT}}$
\begin{equation}
    \tilde{C}_\mathrm{LIFT} = \frac{\tilde{F}_\mathrm{LIFT}}{\Delta \tilde{\dot{X}}} = \tilde{L} \frac{\tilde{\varrho}}{2}  \frac{\tilde{R}_\mathrm{STAT}^{2} + \tilde{R}^{2}}{\tilde{R}_\mathrm{STAT}^{2} - \tilde{R}^{2}} \tilde{\Gamma} = \tilde{\varrho} \tilde{\Omega} \tilde{V} \frac{\tilde{R}_\mathrm{STAT}^{2} + \tilde{R}^{2}}{\tilde{R}_\mathrm{STAT}^{2} - \tilde{R}^{2}}.
\end{equation}

To calculate the added mass $\tilde{M}_{\mathrm{VIRT}}$ and the cross-coupled damping $\tilde{C}_{\mathrm{LIFT}}$ of the test rig, its geometry is approximated by a number of $N$ hollow cylindrical disks with the corresponding radii $\tilde{R}_{i}$ and $R_{i,\mathrm{STAT}}$ and volumes $\tilde{V}_i$.
\begin{equation}
    \tilde{M}_{\mathrm{VIRT}} = \tilde{\varrho} \sum \limits_{i}^{N} \tilde{V}_i \frac{\tilde{R}_{i,\mathrm{STAT}}^2 + \tilde{R}_{i}^2}{\tilde{R}_{i,\mathrm{STAT}}^2 - \tilde{R}_{i}^2}, \quad
    \tilde{C}_\mathrm{LIFT} = \tilde{\varrho} \tilde{\Omega} \sum \limits_{i}^{N} \tilde{V}_i \frac{\tilde{R}_{i,\mathrm{STAT}}^2 + \tilde{R}_{i}^2}{\tilde{R}_{i,\mathrm{STAT}}^2 - \tilde{R}_{i}^2}.
\end{equation}

The added mass and cross-coupled damping of the test rig at an angular frequency $\tilde{\Omega} = 138.2 \, \mathrm{1/s}$ yields $\tilde{M}_{\mathrm{VIRT}} = 19.0 \, \mathrm{kg}$ and $\tilde{C}_\mathrm{LIFT} = 2. 61 \times 10^3 \, \mathrm{Ns / m}$ . 
Together with the weighted mass of the rotor $\tilde{M}_{\mathrm{ROT}}$, the added mass $\tilde{M}_{\mathrm{VIRT}}$ forms an effective mass $\tilde{M}_{\mathrm{EFF}} = \tilde{M}_{\mathrm{VIRT}} + \tilde{M}_{\mathrm{ROT}}$, which can be determined experimentally.
\begin{figure}
	\centering
	\includegraphics[scale=1.0]{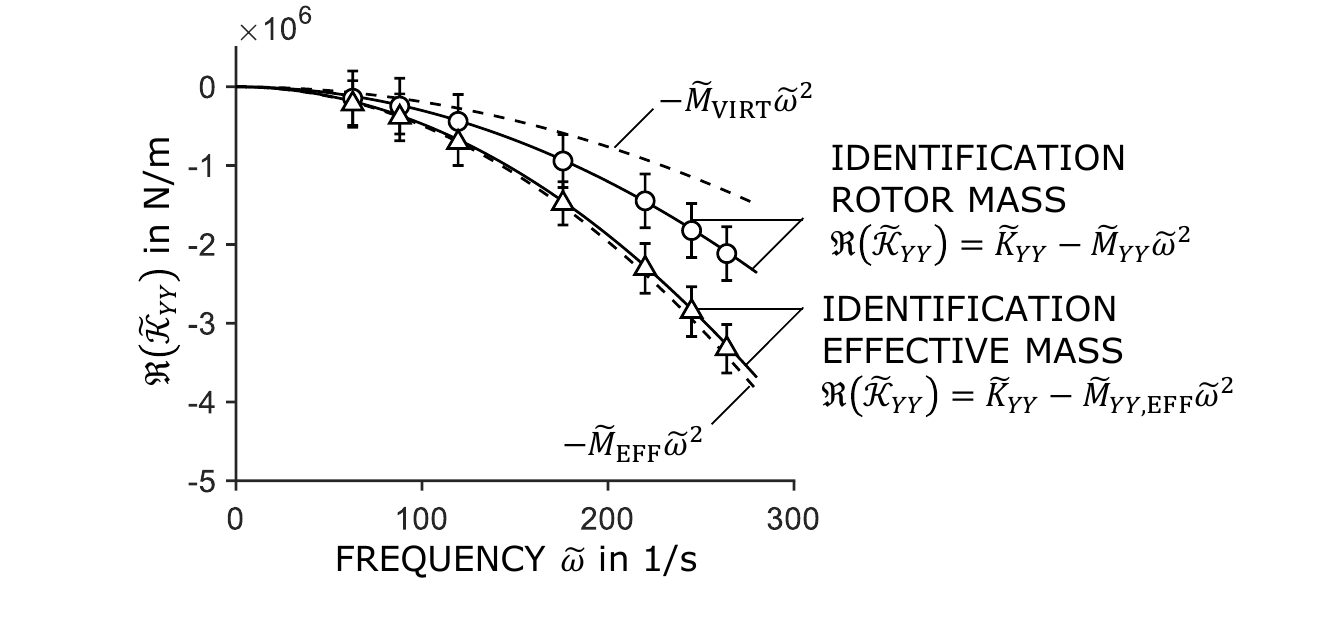}
	\caption{Identification of the effective mass of the annular gap module $\tilde{M}_{\mathrm{EFF}}$ consisting of the mass of the rotor $\tilde{M}_{\mathrm{ROT}}$ and the added mass in the annular gap module $\tilde{M}_{\mathrm{VIRT}}$.}
	\label{fig:complex_stiffness_added_mass}
\end{figure}

Figure~\ref{fig:complex_stiffness_added_mass} shows the real part of the complex stiffness $\Re ( \tilde{\mathcal{K}}_{YY} )$ in analogy to the validation of the identification procedure. Here, the markers represent the experimentally identified mass of the rotor $\tilde{M}_{YY}$, as well as the effective mass of the gap flow test rig~$\tilde{M}_{YY,\mathrm{EFF}}$. Again, the solid lines represent the regression curves, whereas the dashed lines show the calculated complex stiffnesses $-\tilde{M}_{\mathrm{VIRT}} \, \tilde{\omega}^2$ and $-\tilde{M}_{\mathrm{EFF}} \, \tilde{\omega}^2$. It is shown that the identified effective mass~$\tilde{M}_{YY,\mathrm{EFF}}$ is in good agreement with the calculated one~$\tilde{M}_{\mathrm{EFF}}$. Compared to the identification of the rotor mass, cf. figure~\ref{fig:complex_stiffness_repeatability}, the increased deviation, i.e. $6.8\,\%$, results from the assumptions of the potential-theoretical calculation of the added mass $\tilde{M}_{\mathrm{VIRT}}$, as well as the approximation of the test rig geometry by cylindrical disks.

\begin{figure}
	\centering
	\includegraphics[scale=1.0]{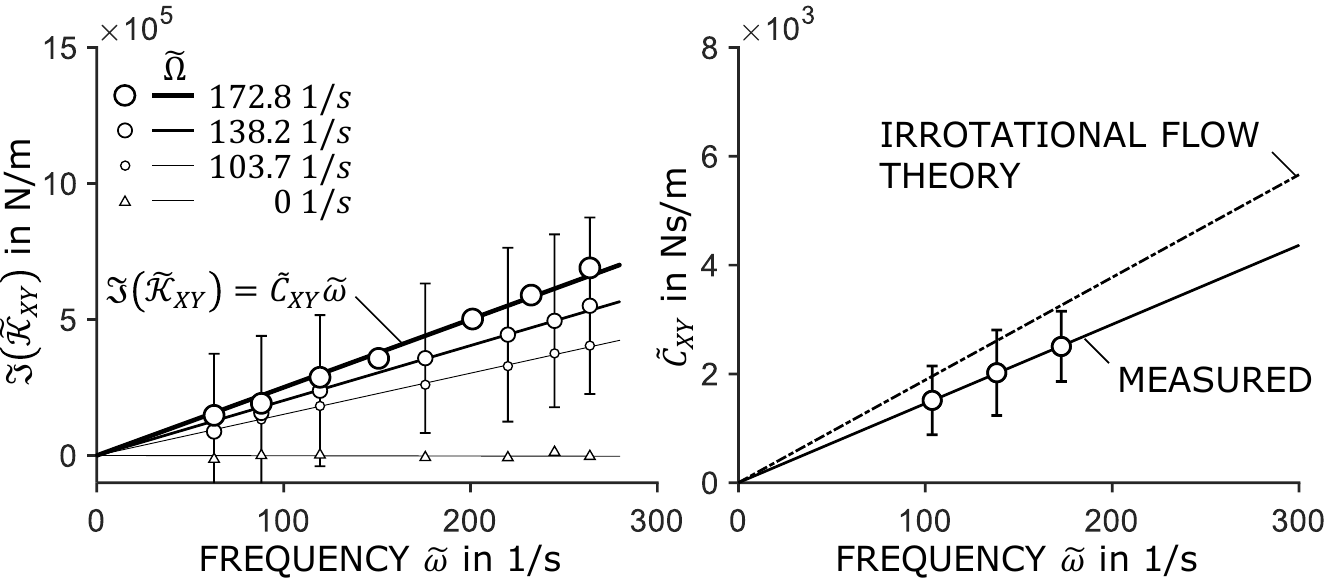}
	\caption{Identification of the cross-coupled damping $\tilde{C}_{XY}$ of the annular gap module by translational excitation (left) and comparison with the predictions by irrotational flow theory (right).}
	\label{fig:complex_stiffness_added_lift}
\end{figure}

Subsequently, the identification of the cross-coupled damping~$\tilde{C}_{XY}$, i.e. the imaginary part of the complex stiffness $\Im(\tilde{\mathcal{K}}_{XY})$, is shown on the left-hand side in figure~\ref{fig:complex_stiffness_added_lift}. 
To identify the predicted influence of the angular frequency of the rotor $\tilde{\Omega}$ on the cross-coupled damping~$\tilde{C}_{XY}$, it is varied in three steps. The marker size as well as the line width correlates with the angular frequency of the rotor $\tilde{\Omega}$. Larger markers and thicker lines represent larger angular frequencies of the rotor. For improved readability, error bars are only given for an angular frequency of $\tilde{\Omega} = 138.2\,\mathrm{1/s}$. However, the measurement uncertainty is representative for all measurements. It can be seen, that the imaginary part of the complex stiffness $\Im(\tilde{\mathcal{K}}_{XY})$ increases linearly with the angular frequency of the rotor $\tilde{\Omega}$ as well as with the precessional frequency $\tilde{\omega}$. The identified cross-coupled damping of the test rig $\tilde{C}_{XY}$ is in the same order of magnitude as the cross-coupled damping of an annular gap, i.e. $\tilde{C}_{XY} \sim 10^3 \, \mathrm{Ns/m}$. 

Subsequently, the right-hand side of figure~\ref{fig:complex_stiffness_added_lift} shows the comparison of the identified cross-coupled damping~$\tilde{C}_{XY}$, i.e. the slopes of the straight lines on the left-hand side of figure~\ref{fig:complex_stiffness_added_lift}, for the three angular frequencies of the rotor in comparison with the theoretical calculations based on irrotational flow theory. In comparison to the calculations, it can be seen that the identified cross-coupled damping is below the calculated one, resulting in a deviation of $22\,\%$. Here, the deviations can be attributed on the one hand to the already mentioned approximation of the test rig geometry by hollow cylindrical disks and on the other hand to the assumptions of the irrotational flow theory calculations. It should be noted, that the given potential of the flow, cf. equation~\ref{eqn:potential_flow_annular_gap}, only fulfils the boundary condition at the rotor. However, the boundary condition at the stator surface, i.e. $\tilde{c}_\varphi |_{r = R_{\mathrm{STAT}}} = 0$, is violated. Nevertheless, in comparison with the experimentally identified cross-coupled damping~$\tilde{C}_{XY}$, it is shown that the given potential of the flow is a decent approximation.

\section{Experimental and simulation results}\label{sec:results}
\subsection{Tested annuli and test conditions}
The experimental investigations are performed using three different annuli with lengths $\tilde{L} = 71.5\,\mathrm{mm}$, $92.95\,\mathrm{mm}$ and $114.4\,\mathrm{mm}$. The three lengths correspond to the dimensionless lengths ${L = 1.0;\,1.3;\,1.6}$. The selection of the range is primarily based on the desired investigation of the additional rotordynamic coefficients, which, according to the literature, only become relevant from a gap length of $L \geq L_{\mathrm{CRIT}} = 1.5$, cf. \citet{Childs.1993}. All experiments are conducted using water at a temperature of $35 \, ^\circ \mathrm{C}$ over an eccentricity range $0.1 \le \varepsilon \le 0.8$. To prevent cavitation within the annulus, the base pressure within the test rig is chosen to be nine times the ambient pressure, i.e. $9 \, \mathrm{bar}$. The flow number $\phi$ as well as the pre-swirl ratio $C_\varphi |_{z=0}$, the centre of rotation $z_T$ and the modified Reynolds number $Re_\varphi^*:=\psi Re_\varphi^{n_\mathrm{f}}$ are kept constant. Their influence will be the subject of future experiments. Following the results of \citet{Kuhr.2022b}, the flow number is $\phi = 0.7$, whereas the pre-swirl, the modified Reynolds number and the centre of rotation are $C_\varphi |_{z=0}$, $z_T=0.5$ and $Re_\varphi^* = 0.031$. The modified Reynolds number $Re_\varphi^*$ corresponds to a Reynolds number of $Re_\varphi = 4000$, indicating fully turbulent flow inside the annulus, cf. \citet{Zirkelback.1996}, an empirical exponent $n_\mathrm{f} = 0.24$ and a relative gap clearance of $\psi = 4.2 \, \permil$.  
The experiments are compared to the Clearance-Averaged Pressure Model (CAPM) by \citet{Kuhr.2022c}. The simulation method uses a perturbed integro-differential approach, similar to the work of \citet{Childs.1982,Simon.1992b} and \citet{SanAndres.1993c}, with a Hirs' model for the wall shear stresses and power-law ansatz functions for the velocity profiles, to calculate the dynamic force and moment characteristics of the annulus. For the comparison with the experimental results, a fully turbulent flow with in annulus is assumed. Accordingly, the empirical constants of the Hirs' wall friction model were set to $m_\mathrm{f} = 0.0645$ and $n_\mathrm{f} = 0.24$. The inlet pressure loss coefficient was set to $\zeta = 0.25$.

Figures~\ref{fig:stiffness_Lxx_z005_Re4000_Ga05_Phi07} to~\ref{fig:inertia_Lxx_z005_Re4000_Ga05_Phi07} show the experimental results in comparison with the calculation method for the stiffness~$K$, damping~$C$ and inertia $M$ coefficients. The individual diagrams in the figures are arranged according to the equation of motion. The first column in the figures represents the rotordynamic coefficients due to translational motion and the second column the coefficients due to angular motion. The rows show the forces and moments on the rotor. The first and second rows show the coefficients derived from the forces $F_X$ and $F_Y$, whereas the third and fourth rows show the coefficients from the moments $M_X$ and $M_Y$. Accordingly, two diagrams arranged above each other represent a sub-matrix. Within the figures, the markers represent the experimentally identified coefficients and the lines represent the calculation results of the CAPM. The marker sizes as well as the line widths correlate with the length of the annular gap. The larger the symbol or the greater the line thickness, the longer the annular gap length. The circular markers as well as the solid lines represent the direct coefficients, whereas the triangular markers and the dashed lines show the cross-coupled coefficients. For improved readability, the experimentally identified coefficients are only provided with error bars at every second measuring point and exclusively on an annulus length of $L = 1.3$. The given measurement uncertainty is representative for all measurements.

\subsection{Stiffness coefficients}
\begin{figure}
	\centering
	\includegraphics[scale=1.0]{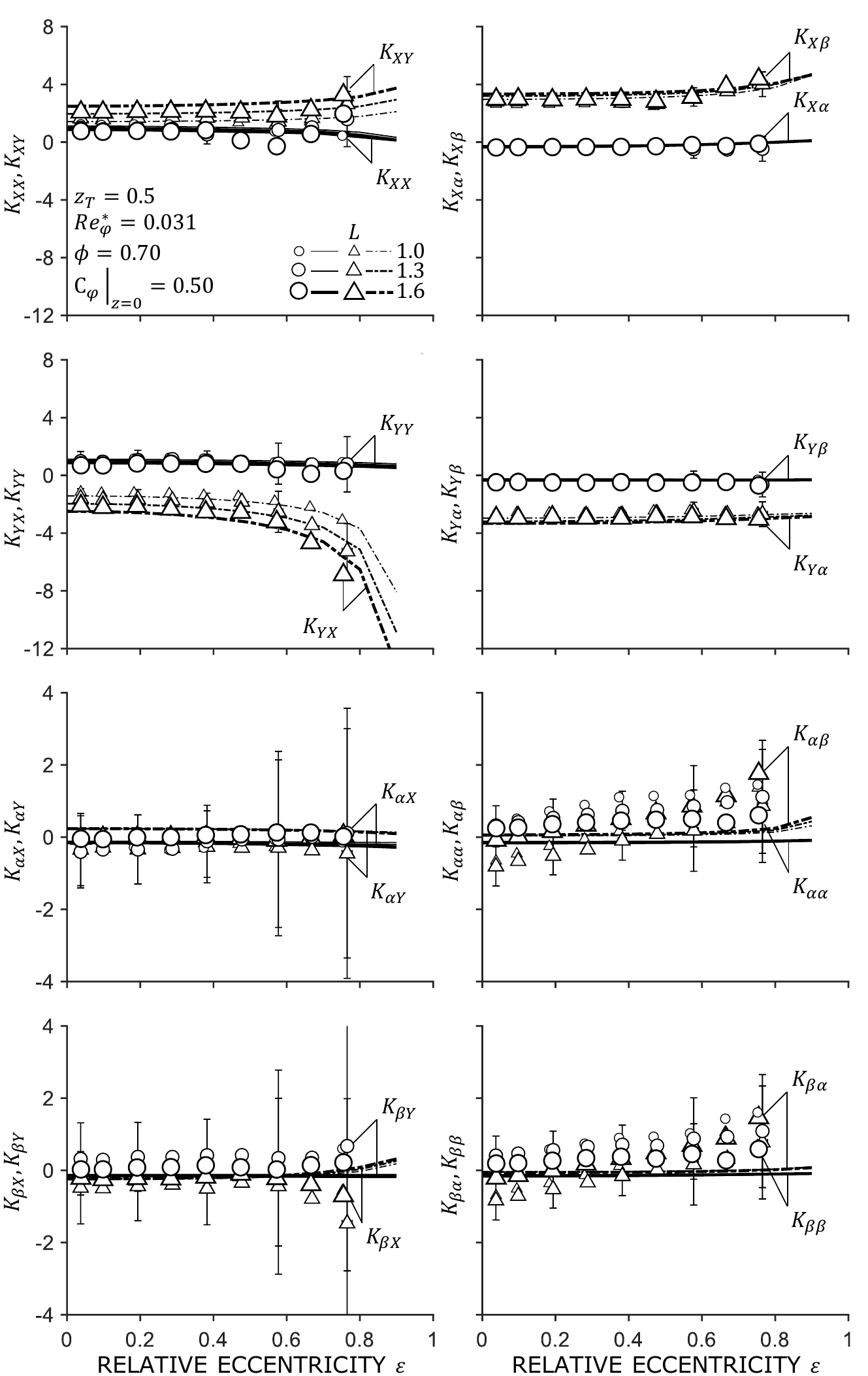}
	\caption{Experimental and simulation results for the stiffness coefficients varying the annulus length.}
	\label{fig:stiffness_Lxx_z005_Re4000_Ga05_Phi07}
\end{figure}
Figure~\ref{fig:stiffness_Lxx_z005_Re4000_Ga05_Phi07} shows the experimental and simulation results for the stiffness coefficients~$K$ when varying the annular gap length~$L$. First, the results for the sub-matrices~I and II, i.e. the stiffness coefficients from the forces due to translational and angular motions, are examined. The experimentally identified stiffness coefficients as well as the CAPM predicted stiffness coefficients agree well for all investigated annular gap lengths. There is both a qualitative and a quantitative agreement of the coefficients. All calculated rotordynamic coefficients are within the measurement uncertainty of the experimental results. The variation of the annular gap length~$L$ is mainly shown in the cross-coupled stiffness coefficients $K_{XY}$ and $K_{YX}$ or $K_{x \beta}$ and $K_{X \alpha}$. The dependence of the cross-coupled stiffness coefficients $K_{XY}$ and $K_{YX}$ on the annular gap length $L$ is greater than the dependence of the secondary stiffness coefficients $K_{x \beta}$ and $K_{X \alpha}$. It should be noted that the stiffness coefficients of the sub-matrix~II~$\boldsymbol{K}_\mathrm{II}$ are of the same order of magnitude as the stiffness coefficients of the sub-matrix~I~$\boldsymbol{K}_\mathrm{I}$. This is interesting insofar as with the annular gap lengths of $L = 1.0$ and $L = 1.3$ the typical threshold value of $L_\mathrm{CRIT} = 1.5$ given in the literature for neglecting the additional 36 rotordynamic coefficients, cf. \citet{Childs.1993}, is already subceeded. Furthermore, it is shown that the stiffness coefficients of sub-matrix~I~$\boldsymbol{K}_\mathrm{I}$ are skew-symmetric up to a relative eccentricity of approx. $\varepsilon \leq 0.4$, i.e. $\boldsymbol{K}_\mathrm{I} = -\boldsymbol{K}_\mathrm{I}^T$. This behaviour is representative at all stiffness, damping and inertia coefficients and confirms the value of $\varepsilon \leq 0.5$ given in the literature, cf. \cite{Childs.1993, Gulich.2010}. However, the coefficients of sub-matrix~II~$\boldsymbol{K}_\mathrm{II}$ are skew-symmetric up to a relative eccentricity of $\varepsilon \leq 0.6$.

Subsequently, the results for sub-matrices~III and IV, i.e. the stiffness coefficients from the moments due to translational and angular motion, are examined. Foremost, it is noticeable that the stiffness coefficients of sub-matrices~III~and IV $\boldsymbol{K}_\mathrm{III}$ and $\boldsymbol{K}_\mathrm{IV}$ are one to two orders of magnitude smaller than the stiffness coefficients of sub-matrices~I and II~$\boldsymbol{K}_\mathrm{I}$ and $\boldsymbol{K}_\mathrm{II}$. In addition, an increased measurement uncertainty of the coefficients is shown. This is due to the calculation of the induced moments~$M_X$~and~$M_Y$ as well as the increased measurement uncertainty due to the reduction of the excitation amplitude at eccentricities larger than $\varepsilon \geq 0.5$. The moments $M_X$ and $M_Y$ are not measured directly. Rather, they result from the measured forces $F_X$ and $F_Y$ with the geometric dimensions of the test rig, i.e. the distance of the location of the force measurement in the magnetic bearings to the location of the centre of rotation~$\tilde{z}_T$ of the rotor. Accordingly, for the calculation of the measurement uncertainty of the induced moments~$\delta (M_X)$~and~$\delta (M_Y)$, the measurement uncertainty of the geometric dimension is also considered. Here, the measurement uncertainty is two orders of magnitude larger than the measurement uncertainty of the annular gap length~$\delta (L)$.

The stiffness coefficients of sub-matrices~III~$\boldsymbol{K}_\mathrm{III}$ agree well with the model predictions. At high eccentricities, i.e. $\varepsilon > 0.7$, small differences between the experimental and simulation results become apparent. For example, the identified stiffness coefficients $K_{\beta Y}$ and $K_{\beta X}$ show an opposite behaviour compared to the predicted coefficients by the model. 

Subsequently, the results of the stiffness coefficients of sub-matrix~IV $\boldsymbol{K}_\mathrm{IV}$ are examined. Here, the largest discrepancies between the model prediction and the experimentally identified coefficients are shown. While the calculated coefficients are almost independent of the relative eccentricity $\varepsilon$, the experimentally determined stiffness coefficients of sub-matrix~IV $\boldsymbol{K}_\mathrm{IV}$ increase with relative eccentricity $\varepsilon$. Here, the stiffness coefficients of the sub-matrix~IV $\boldsymbol{K}_\mathrm{IV}$ originate from the induced moments on the rotor $M_X$ and $M_Y$ and the angular degrees of freedom $\alpha$ and $\beta$. Accordingly, only the measured moments and the measured angles are decisive for a deviation from the model predictions. The induced moments on the rotor $M_X$ and $M_Y$ largely dependent on the geometry of the annulus. Even small deviations of the shape from an ideal cylindrical geometry, as it is assumed by the calculation method CAPM, lead to significant changes in the induced moments on the rotor $M_X$ and $M_Y$. Due to the manufacturing tolerances of the annulus, such deviations are immanent within the experimental identification. However, this hypothesis could not be confirmed conclusively. Nevertheless, all model predictions are within the uncertainty bound of the determined coefficients.

In summary, the following statements can be made regarding the stiffness coefficients:
\begin{enumerate}
    \item The length of the annulus significantly influences the coefficients of the first sub-matrix. A dependence of the additional coefficients on the length is recognisable, but less pronounced. A similar observation can be made when considering the influence of the relative eccentricity. Here, the coefficients of the first sub-matrix are also the ones which are most responsive to variations in the eccentricity.
    \item Even for the shortest ring, i.e. $L=1$, the stiffness coefficients of sub-matrix II are of the same order of magnitude as those of sub-matrix I, indicating that the additional coefficients become relevant much earlier than assumed in the literature. This is in accordance with recent results by \citet{Kuhr.2022, Kuhr.2022c}. The Authors claim that besides the annulus length, as it is highlighted in the literature, the flow number $\phi$ and the modified Reynolds number $Re_\varphi^*$ are crucial when determining the relevance of the additional coefficients.
    \item The stiffness coefficients due to the moments on the rotor are one order of magnitudes smaller than the coefficients of sub-matrices I and II. Therefore, the whole stiffness matrix is dominated by those to sub-matrices.
\end{enumerate}

\clearpage
\subsection{Damping coefficients}
\begin{figure}
	\centering
	\includegraphics[scale=1.0]{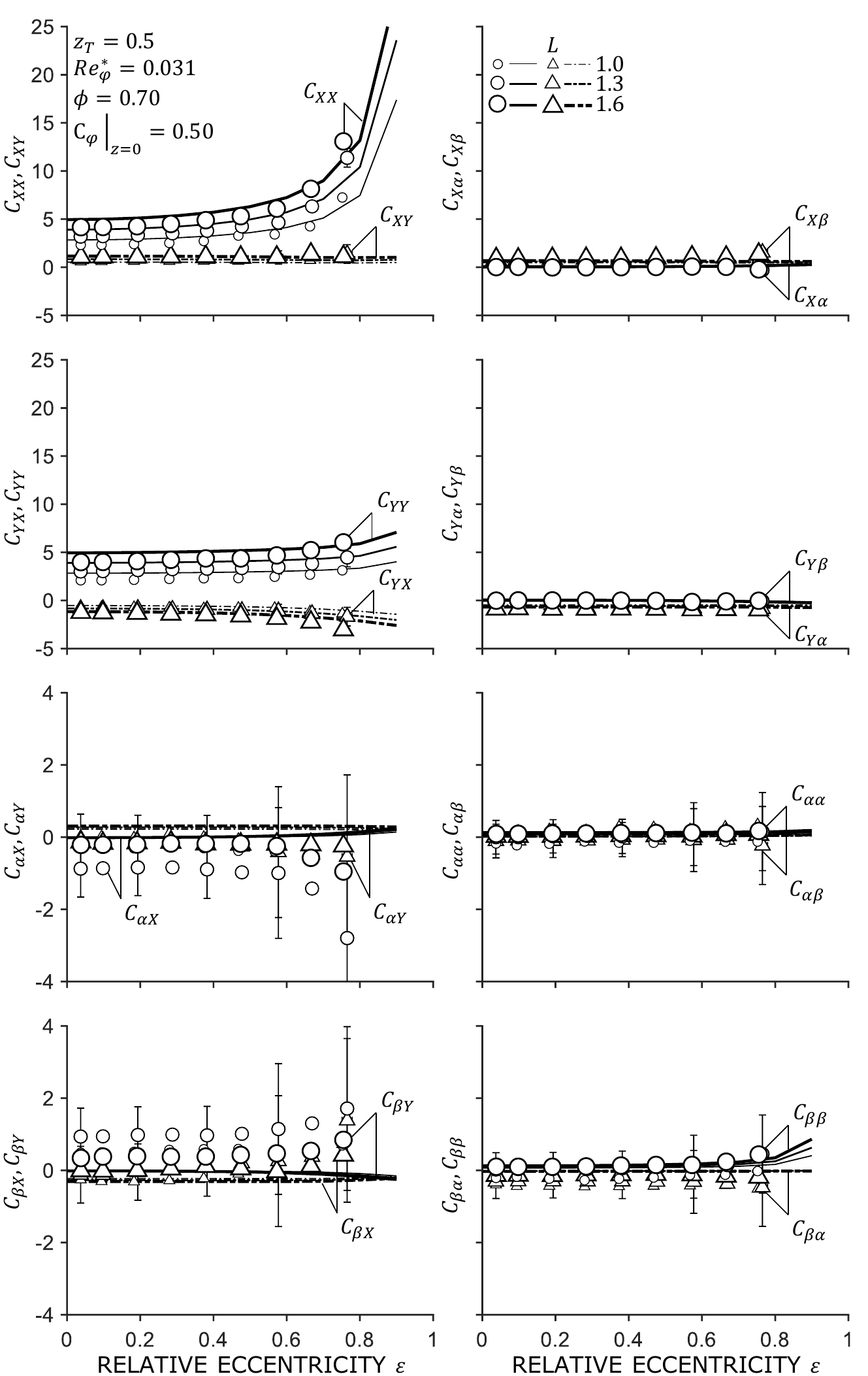}
	\caption{Experimental and simulation results for the damping coefficients varying the annulus length.}
	\label{fig:damping_Lxx_z005_Re4000_Ga05_Phi07}
\end{figure}
Figure~\ref{fig:damping_Lxx_z005_Re4000_Ga05_Phi07} shows the experimental and simulation results for the damping coefficients~$C$. In analogy to the stiffness coefficients, first, the results of sub-matrices~I and II, i.e. the damping coefficients from the forces due to translational and angular motions, are examined. The experimentally identified damping coefficients due to translational and angular motions $\boldsymbol{C}_\mathrm{I}$ and $\boldsymbol{C}_\mathrm{II}$ are in good agreement with the model predictions. It should be noted that in contrast to the stiffness coefficients, the damping coefficients of sub-matrix~II~$\boldsymbol{C}_\mathrm{II}$ are not of the same order of magnitude as the ones of sub-matrix~I~$\boldsymbol{C}_\mathrm{I}$. Rather, the damping coefficients of sub-matrix~II~$\boldsymbol{C}_\mathrm{II}$ are an order of magnitude smaller than the coefficients of sub-matrix~I~$\boldsymbol{C}_\mathrm{I}$. Within sub-matrix~I~$\boldsymbol{C}_\mathrm{I}$, the direct damping coefficients~$C_{XX}$ and $C_{YY}$ are dominant. Here, a clear dependency on the annulus length~$L$ is evident. The cross-coupled damping coefficients~$C_{XY}$ and $C_{YX}$ only slightly dependent on the annular gap length~$L$. A correlation between the annular gap length~$L$ and the damping coefficients of the sub-matrix~II~$\boldsymbol{C}_\mathrm{II}$ cannot be seen in the experiments.

Subsequently, the results of sub-matrices~III and IV, i.e. the damping coefficients from the moments due to translational and angular motions $\boldsymbol{C}_\mathrm{III}$ and $\boldsymbol{C}_\mathrm{IV}$, are examined. Here, the damping coefficients of sub-matrices~III and IV $\boldsymbol{C}_\mathrm{III}$ and $\boldsymbol{C}_\mathrm{IV}$ are an order of magnitude smaller than the damping coefficients of sub-matrices~I and II~$\boldsymbol{C}_\mathrm{I}$ and $\boldsymbol{C}_\mathrm{II}$. The largest deviations between the experimental and simulation results can be seen in sub-matrix~III~$\boldsymbol{C}_\mathrm{III}$. Here, the calculations overestimates the damping coefficients~$C_{\alpha X}$ and $C_{\alpha Y}$, whereas the damping coefficients~$C_{\beta Y}$ and $C_{\beta X}$ are underestimated. However, all calculated coefficients are within the measurement uncertainty. 

In summary, the following statements can be made regarding the damping coefficients:
\begin{enumerate}
    \item The length of the annulus significantly influences the coefficients of the first sub-matrix. A dependence of the additional coefficients on the length is recognisable, but way less pronounced. A similar observation can be made when considering the influence of the relative eccentricity. Here, the coefficients of the first sub-matrix are also the ones which are most responsive to variations in the eccentricity.
    \item The Clearance-Averaged Pressure Model by \citet{Kuhr.2022c} overestimates the damping coefficients~$C_{\alpha X}$ and $C_{\alpha Y}$, whereas the damping coefficients~$C_{\beta Y}$ and $C_{\beta X}$ are underestimated. This is mainly due to the correction of the identified damping coefficients due to the lift generated by an excited rotor within an annulus.
    \item The damping coefficients of sub-matrix II, III and IV are one to two orders of magnitudes smaller than the damping coefficients of sub-matrix I. Therefore, the whole damping matrix is dominated by sub-matrix I.
\end{enumerate}

\clearpage
\subsection{Inertia coefficients}
\begin{figure}
	\centering
	\includegraphics[scale=1.0]{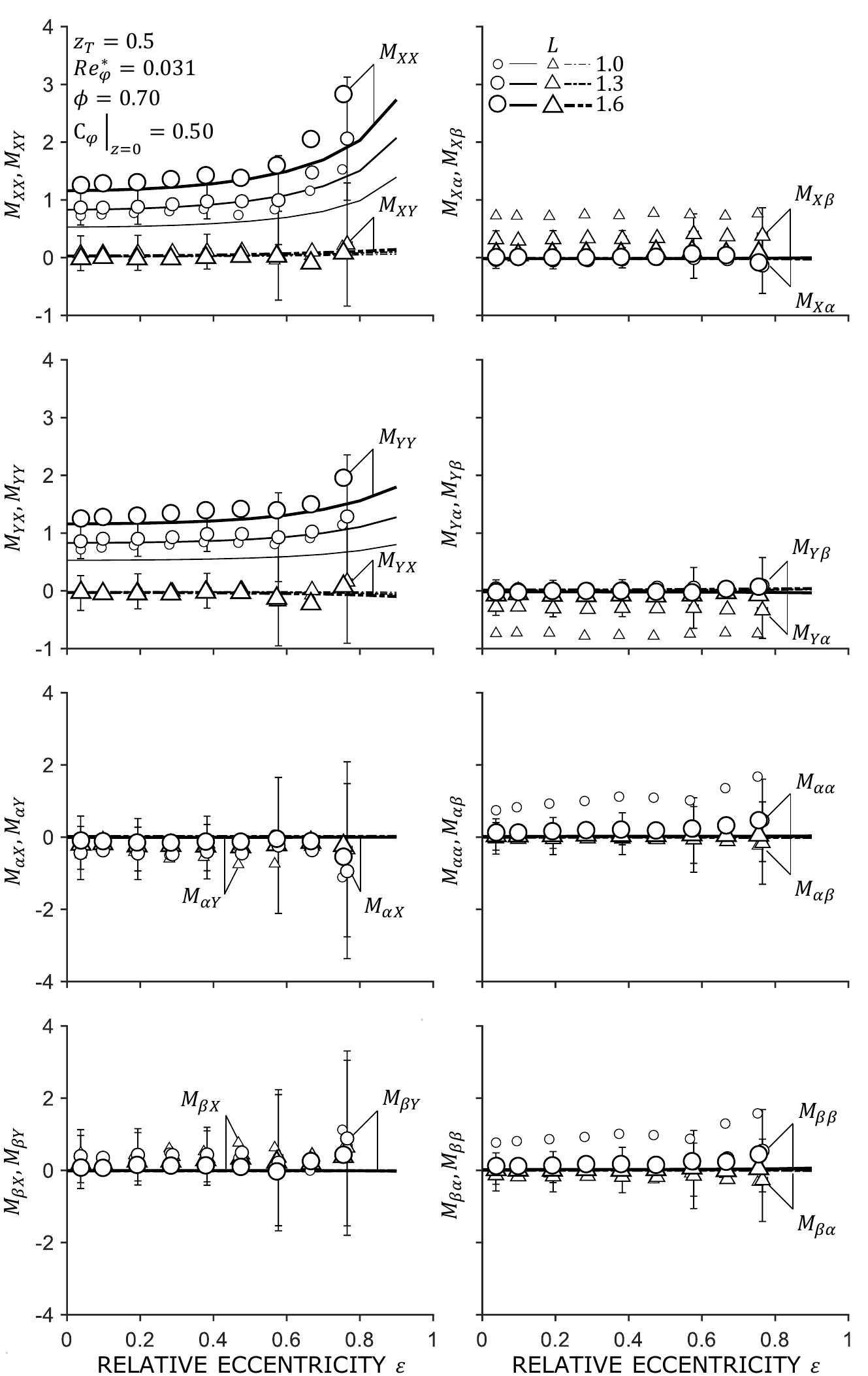}
	\caption{Experimental and simulation results for the inertia coefficients varying the annulus length.}
	\label{fig:inertia_Lxx_z005_Re4000_Ga05_Phi07}
\end{figure}
At last, figure~\ref{fig:inertia_Lxx_z005_Re4000_Ga05_Phi07} shows the experimental and simulation results for the inertia coefficients $M$. First, the results of sub-matrices~I and II, i.e. the inertia coefficients from the forces due to translational and angular motions $\boldsymbol{M}_\mathrm{I}$ and $\boldsymbol{M}_\mathrm{II}$, are examined. The identified inertia coefficients of the sub-matrix~I~$\boldsymbol{M}_\mathrm{I}$ agree well with the model predictions of the clearance-averaged pressure model up to a relative eccentricity of $\varepsilon \leq 0.6$. For larger relative eccentricities, the model underestimates the experimentally identified direct inertia coefficients $M_{XX}$ and $M_{YY}$.
Moreover, the direct inertia coefficients $M_{XX}$ and $M_{YY}$ of the shortest annular gap length $L = 1.0$ show a more pronounced deviation over the whole range of relative eccentricities than the inertia coefficients at the annular gap lengths~$L = 1.3$ and $L = 1.6$. 
Here, one explanation for the discrepancy between the identification and the model prediction is the correction of the measurement results based on the preliminary tests and the irrotational flow theory calculations, cf. section~\ref{sec:characteristics_test_rig}. Here, the preliminary tests are carried out without a built-in annular gap sleeve on the stator side. With the annular gap installed, the experimentally determined added mass of the annular gap module $\tilde{M}_{\mathrm{VIRT}}$ must be corrected by the added mass of the stator-side annular gap sleeve. This correction of the dynamic properties of the experimental setup is carried out by means of the irrotational flow theory calculation. However, these calculations are subject to the assumption of a plane flow, i.e. $\tilde{L} / \tilde{R}_{\mathrm{STAT}} \gg 1$. This assumption is increasingly violated as the annular gap length~$L$ becomes shorter. Accordingly, the calculations are only suitable for correction in a good approximation. The longer the investigated annular gap length~$L$, the smaller the deviations between the experimental results and the calculation method. 

Subsequently, the results of the inertia coefficients of sub-matrix~II~$\boldsymbol{M}_\mathrm{II}$ are examined. Here, it is shown that the inertia coefficients of sub-matrix~II~$\boldsymbol{M}_\mathrm{II}$ are an order of magnitude smaller than the inertia coefficients of sub-matrix~I~$\boldsymbol{M}_\mathrm{I}$. Furthermore, with decreasing annular gap lengths, an increasing deviation of the cross-coupled inertia coefficients $M_{X \beta}$ and $M_{Y \alpha}$ between the experimental and simulation results becomes apparent. This increasing deviation is as well explained by the correction of the preliminary tests. Section~\ref{sec:characteristics_test_rig} only focusses on the added mass due to translational motions of the rotor. From the discipline of marine engineering \cite{Newman.1986,Moreau.2009}, however, it is known that in addition to the added mass due to translational motions $\tilde{M}_{\mathrm{VIRT}} = \tilde{M}_{XX, \mathrm{VIRT}} = \tilde{M}_{YY, \mathrm{VIRT}}$ also the added masses due to angular motions $\tilde{M}_{X\beta ,\mathrm{VIRT}} = - \tilde{M}_{Y\alpha , \mathrm{VIRT}}$ and the added inertia coefficients due to translational and angular motions $\tilde{M}_{\alpha Y ,\mathrm{VIRT}} = -\tilde{M}_{\beta X , \mathrm{VIRT}}$ and $\tilde{\Theta}_{\mathrm{VIRT}} = \tilde{M}_{\alpha\alpha, \mathrm{VIRT}} = \tilde{M}_{\beta\beta, \mathrm{VIRT}}$ must be considered. According to Newman and Moreau \& Korotkin, these additional terms can be obtained by the added mass due to translational motions $\tilde{M}_{\mathrm{VIRT}}$, the geometric dimensions of the test rig and the position of the centre of rotation~$z_T$
\begin{subequations}
    \begin{gather}
        \tilde{M}_{X\beta,\mathrm{VIRT}} = \tilde{M}_{\beta X, \mathrm{VIRT}} = \int\limits_{-\tilde{z}_T}^{\tilde{L}-\tilde{z}_T}  \tilde{z}  \, \frac{ \tilde{M}_\mathrm{VIRT} }{ \tilde{L} } \, \mathrm{d} \tilde{z},\\
        \tilde{M}_{Y \alpha, \mathrm{VIRT}} = \tilde{M}_{\alpha Y, \mathrm{VIRT}} = - \tilde{M}_{X \beta, \mathrm{VIRT}},\\
        \tilde{\Theta}_{\mathrm{VIRT}} =  \int\limits_{-\tilde{z}_T}^{\tilde{L}-\tilde{z}_T}  \tilde{z}^2  \, \frac{ \tilde{M}_\mathrm{VIRT} }{ \tilde{L} } \, \mathrm{d} \tilde{z}.
    \end{gather}
\end{subequations}
Accordingly, the correction of the preliminary tests does not only refer to the main inertia coefficients $M_{XX}$ and $M_{YY}$. Rather, the cross-coupled inertia coefficients of the sub-matrices~II and III $M_{X \beta}$ and $M_{Y \alpha}$ and $M_{\alpha Y}$ and $M_{\beta X}$, respectively, as well as main inertia coefficients of the sub-matrix~IV $M_{\alpha \alpha}$ and $M_{\beta \beta }$ are also corrected.

Subsequently, the results for sub-matrices~III and IV, i.e. the inertia coefficients from the moments due to translational and angular motion $\boldsymbol{M}_\mathrm{III}$ and $\boldsymbol{M}_\mathrm{IV}$, are examined.
Similar to the coefficients of sub-matrix~II~$\boldsymbol{M}_\mathrm{II}$, the inertia coefficients of sub-matrix~III and IV $\boldsymbol{M}_\mathrm{III}$ and $\boldsymbol{M}_\mathrm{IV}$ are one order of magnitude smaller than the inertia coefficients of sub-matrix~I~$\boldsymbol{M}_\mathrm{I}$. Again, the experimental and simulation results are in good agreement. Similarly, the largest deviations between model and experimental identification are visible in the inertia coefficients $M_{\alpha Y}$ and $M_{\beta X}$ or $M_{\alpha \alpha}$ and $M_{\beta \beta }$.

In summary, the following statements can be made regarding the inertia coefficients:
\begin{enumerate}
    \item The annulus length linearly influences the direct inertia coefficients $M_{XX}$ and $M_{YY}$. Furthermore, with increasing relative eccentricity, the direct inertia coefficients of the fist sub-matrix experience an exponential growth.
    \item The deviations between the experimental investigations and the calculations original mainly from the corrections based on irrotational flow theory.
    \item Similar to the damping coefficients, the inertia coefficients of sub-matrix I are the dominant rotordynamic coefficients in the whole inertia matrix. 

\end{enumerate}
\clearpage
\section{Conclusions}
Nowadays, most studies on the dynamic properties of annular gaps focus only on the force characteristics due to translational motions, while the tilt and moment coefficients are less well studied. Therefore, there is hardly any reliable experimental data for the additional coefficients that can be used for validation purpose. To improve this, a test rig first presented by \cite{Kuhr.2022b} is used to experimentally determine the dynamic properties of three annuli of different lengths. By using active magnetic bearings, the rotor is excited with user-defined frequencies and the rotor position and the forces and moments induced by the flow field in the annulus are measured. To obtain accurate and reliable experimental data, extensive preliminary studies are carried out to determine the known characteristics of the test rig rotor and the added mass and inertia imposed by the test rig. Subsequently, an elaborate uncertainty quantification is carried out to quantify the measurement uncertainties. Following the preliminary tests, the force and moment characteristics of three annular gaps at three different lengths $L = 1, \, 1.3, \, 1.6$ are identified and compared to a calculation method developed by \citet{Kuhr.2022, Kuhr.2022c}. It is shown that the presented experimental data agree well with the calculation method, especially for the additional rotordynamic tilt and moment coefficients. Furthermore, it is shown that the annulus length significantly influences mainly the coefficients of the first sub-matrix. A dependence of the additional coefficients on the length is recognisable, but less pronounced. Within the damping and inertia matrix the first sub-matrix I, i.e. the coefficients originating from the forces and the translational motion, are dominating. However, regarding the stiffness matrix, the first and second sub-matrix I and II are the dominant ones. Even for the shortest ring $L = 1$ being significantly shorter than the critical length given in the literature $L_\mathrm{CRIT} = 1.5$, the stiffness coefficients of sub-matrix II are of the same order of magnitude as those of sub-matrix I, indicating that the additional coefficients become relevant much earlier than assumed in the literature, cf. \citet{Childs.1993}. This supports the simulation results by \citet{Kuhr.2022, Kuhr.2022c}, indicating that the relevance of the additional coefficients is not solely depending on the annulus length.

\section{Aknowledgements}
The financial support of the Federal Ministry for Economic Affairs and Energy (BMWi) due to an enactment of the German Bundestag under Grant No. 03EE5036B and KSB SE \& Co. KGaA is gratefully acknowledged. In addition, the financial support of the industrial collective research programme (IGF no. 21029 N/1), supported by the Federal Ministry for Economic Affairs and Energy (BMWi) through the AiF (German Federation of Industrial Research Associations e.V.) due to an enactment of the German Bundestag is gratefully acknowledged. Special gratitude is expressed to Mr. Benjamin Hermann, Mr. Philipp Wetterich and Prof. Dr.-Ing. Peter F. Pelz from the Chair of Fluid Systems at the Technische Universität Darmstadt for their help with the test rig, overall input and outstanding scientific discussions.

\clearpage
\section*{Nomenclature}
\begin{tabbing}
\noindent \textbf{Dimensional variables} \\

$ \tilde{C}_{ij}$  \hspace{20mm} \= damping coefficient due to force and translational motion \\ \>($i,j=X,Y$) [N s/m]\\
$ \tilde{C}_{ij}$ \> damping coefficient due to force and angular motion \\ \>($i=X,Y$; $j=\alpha,\beta$) [N s]\\
$ \tilde{C}_{ij}$ \> damping coefficient due to moment and translational motion \\ \>($i=\alpha,\beta$; $j=X,Y$) [Ns]\\
$ \tilde{C}_{ij}$ \> damping coefficient due to moment and angular motion \\ \>($i,j=\alpha,\beta$) [N m s]\\
$ \tilde{e} $ \> eccentricity [m]\\
$ \tilde{F} $ \> force on the rotor [N]\\
$ \tilde{h} $ \> gap function [m]\\
$ \tilde{ \bar{h} } $ \> mean gap height [m]\\
$ \tilde{K}_{ij}$ \> stiffness coefficient due to force and translational motion \\ \>($i,j=X,Y$) [N/m]\\
$ \tilde{K}_{ij}$ \> stiffness coefficient due to force and angular motion \\ \>($i=X,Y$; $j=\alpha,\beta$) [N]\\
$ \tilde{K}_{ij}$ \> stiffness coefficient due to moment and translational motion \\ \>($i=\alpha,\beta$; $j=X,Y$) [N]\\
$ \tilde{K}_{ij}$ \> stiffness coefficient due to moment and angular motion \\ \>($i,j=\alpha,\beta$) [Nm]\\
$ \tilde{ L } $ \> annulus length [m]\\
$ \tilde{M}$ \> moment on the rotor [Nm]\\
$ \tilde{M}_{ij}$ \> inertia coefficient due to force and translational motion \\ \>($i,j=X,Y$) [kg]\\
$ \tilde{M}_{ij}$ \> inertia coefficient due to force and angular motion \\ \>($i=X,Y$; $j=\alpha,\beta$) [kg m]\\
$ \tilde{M}_{ij}$ \> inertia coefficient due to moment and translational motion \\ \>($i=\alpha,\beta$; $j=X,Y$) [kg m]\\
$ \tilde{M}_{ij}$ \> inertia coefficient due to moment and angular motion \\ \>($i,j=\alpha,\beta$) [kg m²]\\
$ \tilde{ R } $ \> shaft radius [m]\\
$ \tilde{c}_\varphi$ \> velocity in $\varphi$ direction [m/s]\\
$ \tilde{c}_z $ \> velocity in $z$ direction [m/s]\\
$ \tilde{\bar{C}}_\varphi |_{z=0}$ \> swirl velocity at the annulus entrance [m/s]\\
$ \tilde{\bar{C}}_z$ \> mean velocity in $z$ direction [m/s]\\
$\tilde{X}, \tilde{Y}, \tilde{z}$ \> spatial coordinates [m]\\
$\tilde{\eta}$ \> dynamic viscosity [Pa s]\\
$\tilde{\varrho}$ \> density [kg/m³]\\
$\tilde{\nu}$ \> kinematic viscosity [m²/s]\\
$\tilde{\Phi}$ \> potential of the flow in the annular gap  [m²/s]\\
$\tilde{\omega}$ \> angular excitation frequency of the rotor [1/s]\\
$\tilde{\Omega}$ \> angular frequency of the rotor [1/s]\\
\\
\noindent \textbf{Nondimensional variables} \\
{$ C_{ij} := {2\tilde{\bar{h}}\tilde{C}_{ij}}/(\tilde{\varrho}{\Omega}\tilde{R}^3\tilde{L})$ } \hspace{20mm} \= damping coefficient due to force and translational motion \\ \>($i,j=X,Y$) [-]\\
{$ C_{ij} := {2\tilde{\bar{h}}\tilde{C}_{ij}}/(\tilde{\varrho}{\Omega}\tilde{R}^3\tilde{L}^2)$ } \> damping coefficient due to force and angular motion \\ \>($i=X,Y$; $j=\alpha,\beta$) [-]\\
{$ C_{ij} := {2\tilde{\bar{h}}\tilde{C}_{ij}}/(\tilde{\varrho}{\Omega}\tilde{R}^3\tilde{L}^2)$ } \> damping coefficient due to moment and translational motion \\ \>($i=\alpha,\beta$; $j=X,Y$) [-]\\
{$ C_{ij} := {2\tilde{\bar{h}}\tilde{C}_{ij}}/(\tilde{\varrho}\tilde{\Omega}\tilde{R}^3\tilde{L}^3)$ } \> damping coefficient due to moment and angular motion \\ \>($i,j=\alpha,\beta$) [-]\\
{$ {c}_\varphi := \tilde{c}_\varphi / (\tilde{\Omega}\tilde{R})$} \> velocity in $\varphi$ direction [-]\\
{$ {c}_z := \tilde{c}_z/ \tilde{\bar{C}}_z$} \> {velocity in $z$ direction [-]}\\
{$ C_\varphi|_{z=0} := \tilde{\bar{C}}_\varphi |_{z=0} / (\tilde{\Omega}\tilde{R})$} \> pre-swirl at the annulus entrance [-]\\
{$ F :=  2\tilde{F}/(\tilde{\varrho}\tilde{\Omega}^2\tilde{R}^3\tilde{L}) $} \> {force on the rotor [-]}\\
{$ h := \tilde{h}/\tilde{\bar{h}} $} \> {gap function [-]}\\
{$ K_{ij} := {2\tilde{\bar{h}}\tilde{K}_{ij}}/(\tilde{\varrho}{\Omega}^2\tilde{R}^3\tilde{L})$ } \> stiffness coefficient due to force and translational motion \\ \>($i,j=X,Y$) [-]\\
{$ K_{ij} := {2\tilde{\bar{h}}\tilde{K}_{ij}}/(\tilde{\varrho}{\Omega}^2\tilde{R}^3\tilde{L}^2)$ } \> stiffness coefficient due to force and angular motion \\ \>($i=X,Y$; $j=\alpha,\beta$) [-]\\
{$ K_{ij} := {2\tilde{\bar{h}}\tilde{K}_{ij}}/(\tilde{\varrho}{\Omega}^2\tilde{R}^3\tilde{L}^2)$ } \> stiffness coefficient due to moment and translational motion \\ \>($i=\alpha,\beta$; $j=X,Y$) [-]\\
{$ K_{ij} := {2\tilde{\bar{h}}\tilde{K}_{ij}}/(\tilde{\varrho}\tilde{\Omega}^2\tilde{R}^3\tilde{L}^3)$ } \> stiffness coefficient due to moment and angular motion \\ \>($i,j=\alpha,\beta$) [-]\\
{$ L := \tilde{L}/\tilde{R} $} \> {annulus length [-]}\\
{$ M :=  2\tilde{M}/(\tilde{\varrho}\tilde{\Omega}^2\tilde{R}^3\tilde{L}^2) $} \> {moment on the rotor [-]}\\
{$ M_{ij} := {2\tilde{\bar{h}}\tilde{M}_{ij}}/(\tilde{\varrho}\tilde{R}^3\tilde{L})$ } \> inertia coefficient due to force and translational motion \\ \>($i,j=X,Y$) [-]\\
{$ M_{ij} := {2\tilde{\bar{h}}\tilde{M}_{ij}}/(\tilde{\varrho}\tilde{R}^3\tilde{L}^2)$ } \> inertia coefficient due to force and angular motion \\ \>($i=X,Y$; $j=\alpha,\beta$) [-]\\
{$ M_{ij} := {2\tilde{\bar{h}}\tilde{M}_{ij}}/(\tilde{\varrho}\tilde{R}^3\tilde{L}^2)$ } \> inertia coefficient due to moment and translational motion \\ \>($i=\alpha,\beta$; $j=X,Y$) [-]\\
{$ M_{ij} := {2\tilde{\bar{h}}\tilde{M}_{ij}}/(\tilde{\varrho}\tilde{R}^3\tilde{L}^3)$ } \> inertia coefficient due to moment and angular motion \\ \>($i,j=\alpha,\beta$) [-]\\
{$m_\mathrm{f}, \, n_\mathrm{f}$} \> empirical constants within the double logarithmic Moody diagram [-]\\
{$ Re_\varphi := \tilde{\Omega}\tilde{R}\tilde{\bar{h}}/\tilde{\nu} $} \> {Reynolds number [-]}\\
$W$ \> weights in the robust bisquare fitting [-]\\
{$ X, Y := \tilde{X}/\tilde{\bar{h}}, \tilde{Y}/\tilde{\bar{h}}$} \> {spatial coordinates [-]}\\
{$ z := \tilde{z}/\tilde{L} $} \> {axial coordinate [-]}\\
{$ \alpha_X, \beta_Y $} \> {angular coordinates [-]}\\
{$ \alpha, \beta = \tilde{L} \, \alpha_X / \tilde{\bar{h}}, \tilde{L} \, \beta_Y / \tilde{\bar{h}}$} \> {normalised angular coordinates [-]}\\
{$ \varepsilon := \tilde{e}/\tilde{\bar{h}} $} \> {relative eccentricity [-]}\\
{$ \zeta $} \> inlet pressure loss coefficient [-]\\
{$ \varphi $} \> {circumferential coordinate [-]}\\
{$ \phi := \tilde{\bar{C}}_z / (\tilde{\Omega}\tilde{R})$} \> {flow number [-]}\\
{$ \psi := \tilde{\bar{h}}/\tilde{R} $} \> {relative gap clearance [-]}\\
{$ Re_\varphi^* := \psi Re_\varphi^{n_f} $} \> {modified Reynolds number [-]}\\

\end{tabbing}

\bibliographystyle{main-model2-names}

\bibliography{main-refs}

\begin{thebibliography}{43}
\expandafter\ifx\csname natexlab\endcsname\relax\def\natexlab#1{#1}\fi
\providecommand{\url}[1]{\texttt{#1}}
\providecommand{\href}[2]{#2}
\providecommand{\path}[1]{#1}
\providecommand{\DOIprefix}{doi:}
\providecommand{\ArXivprefix}{arXiv:}
\providecommand{\URLprefix}{URL: }
\providecommand{\Pubmedprefix}{pmid:}
\providecommand{\doi}[1]{\href{http://dx.doi.org/#1}{\path{#1}}}
\providecommand{\Pubmed}[1]{\href{pmid:#1}{\path{#1}}}
\providecommand{\bibinfo}[2]{#2}
\ifx\xfnm\relax \def\xfnm[#1]{\unskip,\space#1}\fi
\bibitem[{Batchelor(2012)}]{Batchelor.2012}
\bibinfo{author}{Batchelor, G.K.}, \bibinfo{year}{2012}.
\newblock \bibinfo{title}{An Introduction to Fluid Dynamics}.
\newblock \bibinfo{publisher}{{Cambridge University Press}}.
\newblock \DOIprefix\doi{10.1017/CBO9780511800955}.
\bibitem[{Childs(1982)}]{Childs.1982b}
\bibinfo{author}{Childs, D.W.}, \bibinfo{year}{1982}.
\newblock \bibinfo{title}{Rotordynamic moment coefficients for finite-length
  turbulent seals}.
\newblock \bibinfo{journal}{Proceedings of the IFToMM Conference on
  Rotordynamic Problems in Power Plants} , \bibinfo{pages}{371--378}.
\bibitem[{Childs(1993)}]{Childs.1993}
\bibinfo{author}{Childs, D.W.}, \bibinfo{year}{1993}.
\newblock \bibinfo{title}{Turbomachinery rotordynamics: Phenomena, modeling,
  and analysis}.
\newblock A Wiley Interscience publication, \bibinfo{publisher}{Wiley},
  \bibinfo{address}{New York}.
\bibitem[{Childs and Hale(1994)}]{Childs.1994}
\bibinfo{author}{Childs, D.W.}, \bibinfo{author}{Hale, K.},
  \bibinfo{year}{1994}.
\newblock \bibinfo{title}{A test apparatus and facility to identify the
  rotordynamic coefficients of high-speed hydrostatic bearings}.
\newblock \bibinfo{journal}{Journal of Tribology} \bibinfo{volume}{116},
  \bibinfo{pages}{337--343}.
\newblock \DOIprefix\doi{10.1115/1.2927226}.
\bibitem[{Childs and Kim(1986)}]{Childs.1986c}
\bibinfo{author}{Childs, D.W.}, \bibinfo{author}{Kim, C.H.},
  \bibinfo{year}{1986}.
\newblock \bibinfo{title}{Test results for round-hole-pattern damper seals:
  Optimum configurations and dimensions for maximum net damping}.
\newblock \bibinfo{journal}{Journal of Tribology} \bibinfo{volume}{108},
  \bibinfo{pages}{605}.
\newblock \DOIprefix\doi{10.1115/1.3261277}.
\bibitem[{Childs et~al.(1982)Childs, Nelson, Noyes and Dressman}]{Childs.1982}
\bibinfo{author}{Childs, D.W.}, \bibinfo{author}{Nelson, C.C.},
  \bibinfo{author}{Noyes, T.}, \bibinfo{author}{Dressman, J.B.},
  \bibinfo{year}{1982}.
\newblock \bibinfo{title}{A high-reynolds-number seal test facility: Facility
  description and preliminary test data}.
\newblock \bibinfo{journal}{NASA. Lewis Research Center Rotordyn. Instability
  Probl. in High-Performance Turbomachinery} \URLprefix
  \url{https://ntrs.nasa.gov/citations/19830007370}.
\bibitem[{Feng and Jiang(2017)}]{Feng.2017}
\bibinfo{author}{Feng, H.}, \bibinfo{author}{Jiang, S.}, \bibinfo{year}{2017}.
\newblock \bibinfo{title}{Dynamics of a motorized spindle supported on
  water-lubricated bearings}.
\newblock \bibinfo{journal}{Proceedings of the Institution of Mechanical
  Engineers, Part C: Journal of Mechanical Engineering Science}
  \bibinfo{volume}{231}, \bibinfo{pages}{459--472}.
\newblock \DOIprefix\doi{10.1177/0954406215616653}.
\bibitem[{Gasch et~al.(2002)Gasch, Nordmann and Pf{\"u}tzner}]{Gasch.2002}
\bibinfo{author}{Gasch, R.}, \bibinfo{author}{Nordmann, R.},
  \bibinfo{author}{Pf{\"u}tzner, H.}, \bibinfo{year}{2002}.
\newblock \bibinfo{title}{Rotordynamik}.
\newblock \bibinfo{edition}{2., vollst. neubearb. und erw. aufl.} ed.,
  \bibinfo{publisher}{Springer}, \bibinfo{address}{Berlin u.a.}
\bibitem[{Glienicke(1966)}]{Glienicke.1966}
\bibinfo{author}{Glienicke, J.}, \bibinfo{year}{1966}.
\newblock \bibinfo{title}{Experimental investigation of the stiffness and
  damping coefficients of turbine bearings and their application to instability
  prediction}.
\newblock \bibinfo{journal}{Proceedings of the Institution of Mechanical
  Engineers, Conference Proceedings} \bibinfo{volume}{181},
  \bibinfo{pages}{116--129}.
\newblock \DOIprefix\doi{10.1243/PIME_CONF_1966_181_038_02}.
\bibitem[{G{\"u}lich(2010)}]{Gulich.2010}
\bibinfo{author}{G{\"u}lich, J.F.}, \bibinfo{year}{2010}.
\newblock \bibinfo{title}{Kreiselpumpen: Handbuch f{\"u}r Entwicklung,
  Anlagenplanung und Betrieb}.
\newblock \bibinfo{edition}{3} ed., \bibinfo{publisher}{Springer-Verlag},
  \bibinfo{address}{Berlin, Heidelberg}.
\newblock \DOIprefix\doi{10.1007/978-3-642-05479-2}.
\bibitem[{{Joint Committee for Guides in
  Metrology}(2008)}]{JointCommitteeforGuidesinMetrology.2008}
\bibinfo{author}{{Joint Committee for Guides in Metrology}},
  \bibinfo{year}{2008}.
\newblock \bibinfo{title}{Evaluation of measurement data---guide to the
  expression of uncertainty in measurement}.
\newblock \URLprefix \url{https://www.bipm.org/en/publications/guides}.
\bibitem[{Jolly et~al.(2018)Jolly, Arghir, Bonneau and Hassini}]{Jolly.2018}
\bibinfo{author}{Jolly, P.}, \bibinfo{author}{Arghir, M.},
  \bibinfo{author}{Bonneau, O.}, \bibinfo{author}{Hassini, M.A.},
  \bibinfo{year}{2018}.
\newblock \bibinfo{title}{Experimental and theoretical rotordynamic
  coefficients of smooth and round-hole pattern water-fed annular seals}.
\newblock \bibinfo{journal}{Journal of Engineering for Gas Turbines and Power}
  \bibinfo{volume}{140}.
\newblock \DOIprefix\doi{10.1115/1.4040177}.
\bibitem[{Kanemori and Iwatsubo(1992)}]{Kanemori.1992}
\bibinfo{author}{Kanemori, Y.}, \bibinfo{author}{Iwatsubo, T.},
  \bibinfo{year}{1992}.
\newblock \bibinfo{title}{Experimental study of dynamic fluid forces and
  moments for a long annular seal}.
\newblock \bibinfo{journal}{Journal of Tribology} \bibinfo{volume}{114},
  \bibinfo{pages}{773--778}.
\newblock \DOIprefix\doi{10.1115/1.2920947}.
\bibitem[{Kanemori and Iwatsubo(1994)}]{Kanemori.1994b}
\bibinfo{author}{Kanemori, Y.}, \bibinfo{author}{Iwatsubo, T.},
  \bibinfo{year}{1994}.
\newblock \bibinfo{title}{Forces and moments due to combined motion of conical
  and cylindrical whirls for a long seal}.
\newblock \bibinfo{journal}{Journal of Tribology} \bibinfo{volume}{116},
  \bibinfo{pages}{489--498}.
\newblock \DOIprefix\doi{10.1115/1.2928871}.
\bibitem[{Kim and Palazzolo(2018)}]{Kim.2018}
\bibinfo{author}{Kim, E.}, \bibinfo{author}{Palazzolo, A.B.},
  \bibinfo{year}{2018}.
\newblock \bibinfo{title}{Rotordynamic stability effects of shrouded
  centrifugal impellers with combined whirl and precession}.
\newblock \bibinfo{journal}{Journal of Vibration and Acoustics}
  \bibinfo{volume}{140}, \bibinfo{pages}{021007.1--12}.
\newblock \DOIprefix\doi{10.1115/1.4037958}.
\bibitem[{Kuhr(2022)}]{Kuhr.2022}
\bibinfo{author}{Kuhr, M.}, \bibinfo{year}{2022}.
\newblock \bibinfo{title}{Dynamische Eigenschaften axial durchstr{\"o}mter
  Ringspalte}. volume \bibinfo{volume}{Band 29} of
  \textit{\bibinfo{series}{Forschungsberichte zur Fluidsystemtechnik}}.
\newblock \bibinfo{edition}{[1. auflage]} ed., \bibinfo{publisher}{{Shaker
  Verlag}}, \bibinfo{address}{D{\"u}ren}.
\newblock \DOIprefix\doi{10.2370/9783844086102}.
\bibitem[{Kuhr et~al.(2022a)Kuhr, Lang and Pelz}]{Kuhr.2022b}
\bibinfo{author}{Kuhr, M.M.G.}, \bibinfo{author}{Lang, S.R.},
  \bibinfo{author}{Pelz, P.F.}, \bibinfo{year}{2022}a.
\newblock \bibinfo{title}{Static force characteristic of annular gaps --
  experimental and simulation results}.
\newblock \bibinfo{journal}{Journal of Tribology} \bibinfo{volume}{144},
  \bibinfo{pages}{1--18}.
\newblock \DOIprefix\doi{10.1115/1.4054792}.
\bibitem[{Kuhr et~al.(2022b)Kuhr, Nordmann and Pelz}]{Kuhr.2022c}
\bibinfo{author}{Kuhr, M.M.G.}, \bibinfo{author}{Nordmann, R.},
  \bibinfo{author}{Pelz, P.F.}, \bibinfo{year}{2022}b.
\newblock \bibinfo{title}{Dynamic force and moment characteristics of annular
  gaps - simulation results and evaluation of the relevance of the tilt and
  moment coefficients}.
\newblock \bibinfo{journal}{Journal of Tribology} ,
  \bibinfo{pages}{1--42}\DOIprefix\doi{10.1115/1.4055180}.
\bibitem[{Lund(1974)}]{Lund.1974}
\bibinfo{author}{Lund, J.W.}, \bibinfo{year}{1974}.
\newblock \bibinfo{title}{Stability and damped critical speeds of a flexible
  rotor in fluid-film bearings}.
\newblock \bibinfo{journal}{Journal of Engineering for Industry}
  \bibinfo{volume}{96}, \bibinfo{pages}{509--517}.
\newblock \DOIprefix\doi{10.1115/1.3438358}.
\bibitem[{Marquette et~al.(1997)Marquette, Childs and {San
  Andr{\'e}s}}]{Marquette.1997}
\bibinfo{author}{Marquette, O.R.}, \bibinfo{author}{Childs, D.W.},
  \bibinfo{author}{{San Andr{\'e}s}, L.}, \bibinfo{year}{1997}.
\newblock \bibinfo{title}{Eccentricity effects on the rotordynamic coefficients
  of plain annular seals: Theory versus experiment}.
\newblock \bibinfo{journal}{Journal of Tribology} \bibinfo{volume}{119},
  \bibinfo{pages}{443}.
\newblock \DOIprefix\doi{10.1115/1.2833515}.
\bibitem[{Maslen and Schweitzer(2009)}]{Maslen.2009}
\bibinfo{author}{Maslen, E.H.}, \bibinfo{author}{Schweitzer, G.},
  \bibinfo{year}{2009}.
\newblock \bibinfo{title}{Magnetic Bearings}.
\newblock \bibinfo{publisher}{{Springer Berlin Heidelberg}},
  \bibinfo{address}{Berlin, Heidelberg}.
\newblock \DOIprefix\doi{10.1007/978-3-642-00497-1}.
\bibitem[{Matros et~al.(1995)Matros, Ziegler and Nordmann}]{Matros.1995}
\bibinfo{author}{Matros, M.}, \bibinfo{author}{Ziegler, A.},
  \bibinfo{author}{Nordmann, R.}, \bibinfo{year}{1995}.
\newblock \bibinfo{title}{Fluid structure interactions in annular seals of
  centrifugal pumps}.
\newblock \bibinfo{journal}{Tribology Transactions} \bibinfo{volume}{38},
  \bibinfo{pages}{353--363}.
\newblock \DOIprefix\doi{10.1080/10402009508983415}.
\bibitem[{Moreau and Korotkin(2009)}]{Moreau.2009}
\bibinfo{author}{Moreau, R.}, \bibinfo{author}{Korotkin, A.I.},
  \bibinfo{year}{2009}.
\newblock \bibinfo{title}{Added Masses of Ship Structures}.
  volume~\bibinfo{volume}{88} of \textit{\bibinfo{series}{Fluid Mechanics and
  Its Applications}}.
\newblock \bibinfo{publisher}{{Springer Netherlands}},
  \bibinfo{address}{Dordrecht}.
\newblock \URLprefix
  \url{http://site.ebrary.com/lib/alltitles/docDetail.action?docID=10269498},
  \DOIprefix\doi{10.1007/978-1-4020-9432-3}.
\bibitem[{Moreland et~al.(2018)Moreland, Childs and Bullock}]{Moreland.2018}
\bibinfo{author}{Moreland, J.A.}, \bibinfo{author}{Childs, D.W.},
  \bibinfo{author}{Bullock, J.T.}, \bibinfo{year}{2018}.
\newblock \bibinfo{title}{Measured static and rotordynamic characteristics of a
  smooth-stator/grooved-rotor liquid annular seal}.
\newblock \bibinfo{journal}{Journal of Fluids Engineering}
  \bibinfo{volume}{140}.
\newblock \DOIprefix\doi{10.1115/1.4040762}.
\bibitem[{Neumer(1994)}]{Neumer.1994}
\bibinfo{author}{Neumer, T.}, \bibinfo{year}{1994}.
\newblock \bibinfo{title}{Entwicklung einer Versuchsanlage mit aktiver
  Magnetlagerung zur Parameteridentifikation von Fluid-Struktur-Interaktionen
  in Str{\"o}mungsmaschinen: Zugl.: Kaiserslautern, Univ., Diss}. volume
  \bibinfo{volume}{203} of \textit{\bibinfo{series}{Fortschritt-Berichte VDI
  Reihe 11, Schwingungstechnik}}.
\newblock \bibinfo{publisher}{VDI-Verl.}, \bibinfo{address}{D{\"u}sseldorf}.
\bibitem[{Newman(1986)}]{Newman.1986}
\bibinfo{author}{Newman, J.N.}, \bibinfo{year}{1986}.
\newblock \bibinfo{title}{Marine hydrodynamics}.
\newblock \bibinfo{edition}{5. printing} ed., \bibinfo{publisher}{{MIT Pess}},
  \bibinfo{address}{Cambridge, Mass.}
\bibitem[{Nordmann and Massmann(1984)}]{Nordmann.1984}
\bibinfo{author}{Nordmann, R.}, \bibinfo{author}{Massmann, H.},
  \bibinfo{year}{1984}.
\newblock \bibinfo{title}{Identification of stiffness, damping and mass
  coefficients for annular seals}.
\newblock \bibinfo{journal}{NASA, Lewis Research Center, Rotordynamic
  Instability Problems in High-Performance Turbomachinery} .
\bibitem[{Pinkus(1987)}]{Pinkus.1987}
\bibinfo{author}{Pinkus, O.}, \bibinfo{year}{1987}.
\newblock \bibinfo{title}{The reynolds centennial: A brief history of the
  theory of hydrodynamic lubrication}.
\newblock \bibinfo{journal}{Journal of Tribology} \bibinfo{volume}{109},
  \bibinfo{pages}{2--15}.
\newblock \DOIprefix\doi{10.1115/1.3261322}.
\bibitem[{{San Andr{\'e}s}(1991)}]{SanAndres.1991b}
\bibinfo{author}{{San Andr{\'e}s}, L.}, \bibinfo{year}{1991}.
\newblock \bibinfo{title}{Analysis of variable fluid properties, turbulent
  annular seals}.
\newblock \bibinfo{journal}{Journal of Tribology} \bibinfo{volume}{113},
  \bibinfo{pages}{694--702}.
\newblock \DOIprefix\doi{10.1115/1.2920681}.
\bibitem[{{San Andr{\'e}s}(1993a)}]{SanAndres.1993b}
\bibinfo{author}{{San Andr{\'e}s}, L.}, \bibinfo{year}{1993}a.
\newblock \bibinfo{title}{The effect of journal misalignment on the operation
  of a turbulent flow hydrostatic bearing}.
\newblock \bibinfo{journal}{Journal of Tribology} \bibinfo{volume}{115},
  \bibinfo{pages}{355--363}.
\newblock \DOIprefix\doi{10.1115/1.2921643}.
\bibitem[{{San Andr{\'e}s}(1993b)}]{SanAndres.1993c}
\bibinfo{author}{{San Andr{\'e}s}, L.}, \bibinfo{year}{1993}b.
\newblock \bibinfo{title}{Effect of shaft misalignment on the dynamic force
  response of annular pressure seals}.
\newblock \bibinfo{journal}{Tribology Transactions} \bibinfo{volume}{36},
  \bibinfo{pages}{173--182}.
\newblock \DOIprefix\doi{10.1080/10402009308983146}.
\bibitem[{{San Andr{\'e}s} et~al.(1993){San Andr{\'e}s}, Yang and
  Childs}]{SanAndres.1993}
\bibinfo{author}{{San Andr{\'e}s}, L.}, \bibinfo{author}{Yang, Z.},
  \bibinfo{author}{Childs, D.W.}, \bibinfo{year}{1993}.
\newblock \bibinfo{title}{Thermal effects in cryogenic liquid annular
  seals---part ii: Numerical solution and results}.
\newblock \bibinfo{journal}{Journal of Tribology} \bibinfo{volume}{115},
  \bibinfo{pages}{277--284}.
\newblock \DOIprefix\doi{10.1115/1.2921002}.
\bibitem[{Simon and Fr{\^e}ne(1992)}]{Simon.1992b}
\bibinfo{author}{Simon, F.}, \bibinfo{author}{Fr{\^e}ne, J.},
  \bibinfo{year}{1992}.
\newblock \bibinfo{title}{Analysis for incompressible flow in annular pressure
  seals}.
\newblock \bibinfo{journal}{Journal of Tribology} \bibinfo{volume}{114},
  \bibinfo{pages}{431--438}.
\newblock \DOIprefix\doi{10.1115/1.2920902}.
\bibitem[{Spurk(1995)}]{Spurk.1995}
\bibinfo{author}{Spurk, J.H.}, \bibinfo{year}{1995}.
\newblock \bibinfo{title}{Aufgaben zur Str{\"o}mungslehre}.
\newblock \bibinfo{publisher}{{Springer Berlin Heidelberg}},
  \bibinfo{address}{Berlin, Heidelberg}.
\newblock \DOIprefix\doi{10.1007/978-3-662-05911-1}.
\bibitem[{Spurk and Aksel(2008)}]{Spurk.2008}
\bibinfo{author}{Spurk, J.H.}, \bibinfo{author}{Aksel, N.},
  \bibinfo{year}{2008}.
\newblock \bibinfo{title}{Fluid mechanics}.
\newblock \bibinfo{edition}{2} ed., \bibinfo{publisher}{Springer},
  \bibinfo{address}{Berlin and Heidelberg}.
\newblock \DOIprefix\doi{10.1007/978-3-030-30259-7}.
\bibitem[{Sternlicht and Rieger(1967)}]{Sternlicht.1967}
\bibinfo{author}{Sternlicht, B.}, \bibinfo{author}{Rieger, N.F.},
  \bibinfo{year}{1967}.
\newblock \bibinfo{title}{Rotor stability}.
\newblock \bibinfo{journal}{Proceedings of the Institution of Mechanical
  Engineers, Conference Proceedings} \bibinfo{volume}{182},
  \bibinfo{pages}{82--99}.
\newblock \DOIprefix\doi{10.1243/PIME_CONF_1967_182_010_02}.
\bibitem[{{The MathWorks, Inc.}(2021)}]{TheMathWorksInc..2021}
\bibinfo{author}{{The MathWorks, Inc.}}, \bibinfo{year}{2021}.
\newblock \bibinfo{title}{Least-squares fitting}.
\newblock \URLprefix
  \url{https://de.mathworks.com/help/curvefit/least-squares-fitting.html}.
\bibitem[{Tiwari(2018)}]{Tiwari.2018}
\bibinfo{author}{Tiwari, R.}, \bibinfo{year}{2018}.
\newblock \bibinfo{title}{Rotor Systems: Analysis and identification}.
\newblock \bibinfo{edition}{1} ed., \bibinfo{publisher}{{CRC Press}},
  \bibinfo{address}{Boca Raton, FL and London and New York}.
\newblock \URLprefix \url{https://www.taylorfrancis.com/books/9781351863643},
  \DOIprefix\doi{10.1201/9781315230962}.
\bibitem[{Tiwari et~al.(2004)Tiwari, Lees and Friswell}]{Tiwari.2004}
\bibinfo{author}{Tiwari, R.}, \bibinfo{author}{Lees, A.W.},
  \bibinfo{author}{Friswell, M.I.}, \bibinfo{year}{2004}.
\newblock \bibinfo{title}{Identification of dynamic bearing parameters: A
  review}.
\newblock \bibinfo{journal}{The Shock and Vibration Digest}
  \bibinfo{volume}{36}, \bibinfo{pages}{99--124}.
\newblock \DOIprefix\doi{10.1177/0583102404040173}.
\bibitem[{Tiwari et~al.(2005)Tiwari, Manikandan and Dwivedy}]{Tiwari.2005}
\bibinfo{author}{Tiwari, R.}, \bibinfo{author}{Manikandan, S.},
  \bibinfo{author}{Dwivedy, S.K.}, \bibinfo{year}{2005}.
\newblock \bibinfo{title}{A review of the experimental estimation of the rotor
  dynamic parameters of seals}.
\newblock \bibinfo{journal}{The Shock and Vibration Digest}
  \bibinfo{volume}{37}, \bibinfo{pages}{261--284}.
\newblock \DOIprefix\doi{10.1177/0583102405055442}.
\bibitem[{Zakharov(2010)}]{Zakharov.2010}
\bibinfo{author}{Zakharov, S.M.}, \bibinfo{year}{2010}.
\newblock \bibinfo{title}{Hydrodynamic lubrication research: Current situation
  and future prospects}.
\newblock \bibinfo{journal}{Journal of Friction and Wear} \bibinfo{volume}{31},
  \bibinfo{pages}{56--67}.
\newblock \DOIprefix\doi{10.3103/S106836661001006X}.
\bibitem[{Zeier et~al.(2012)Zeier, Hoffmann and Wollensack}]{Zeier.2012}
\bibinfo{author}{Zeier, M.}, \bibinfo{author}{Hoffmann, J.},
  \bibinfo{author}{Wollensack, M.}, \bibinfo{year}{2012}.
\newblock \bibinfo{title}{Metas.unclib ---a measurement uncertainty calculator
  for advanced problems}.
\newblock \bibinfo{journal}{Metrologia} \bibinfo{volume}{49},
  \bibinfo{pages}{809--815}.
\newblock \DOIprefix\doi{10.1088/0026-1394/49/6/809}.
\bibitem[{Zirkelback and {San Andr{\'e}s}(1996)}]{Zirkelback.1996}
\bibinfo{author}{Zirkelback, N.}, \bibinfo{author}{{San Andr{\'e}s}, L.},
  \bibinfo{year}{1996}.
\newblock \bibinfo{title}{Bulk-flow model for the transition to turbulence
  regime in annular pressure seals}.
\newblock \bibinfo{journal}{Tribology Transactions} \bibinfo{volume}{39},
  \bibinfo{pages}{835--842}.
\newblock \DOIprefix\doi{10.1080/10402009608983602}.

\end{thebibliography}


\end{document}